\documentclass[times, review, 10pt]{elsarticle}

\usepackage[a4paper,top=4.3cm,right=4.8cm,bottom=4.3cm,left=4.8cm]{geometry}
\usepackage{graphicx}
\usepackage{float}
\usepackage{placeins}
\usepackage{booktabs}
\usepackage{multirow}
\usepackage{amsmath}
\usepackage{amssymb}
\usepackage{url}
\usepackage{algorithm}
\usepackage{algorithmic}
\usepackage{rotating}
\usepackage[font=footnotesize]{caption}
\biboptions{numbers,sort&compress}

\begin{document}
\begin{frontmatter}
\title{\fontsize{14}{17}\selectfont NESA: Non-stationary Evidence-aware Semantic Alignment for Robust ECG--Text Representation Pretraining}
\author[inst1]{\fontsize{8}{10}\selectfont Hongjun Liu\fnref{equal}}
\ead{D202210386@xs.ustb.edu.cn}
\author[inst1]{\fontsize{8}{10}\selectfont Rujun Han\fnref{equal}}
\ead{hrj66986698@gmail.com}
\author[inst1]{\fontsize{8}{10}\selectfont Leyu Zhou}
\ead{u202343397@xs.ustb.edu.cn}
\author[inst2]{\fontsize{8}{10}\selectfont Chao Yao\corref{cor1}}
\ead{yaochao@ustb.edu.cn}
\cortext[cor1]{Corresponding author}
\renewcommand{\thefnote}{$\dagger$}
\fntext[equal]{Hongjun Liu and Rujun Han contributed equally to this work.}
\address[inst1]{\fontsize{8}{10}\selectfont School of Artificial Intelligence, University of Science and Technology Beijing, Beijing, China}
\address[inst2]{\fontsize{8}{10}\selectfont School of Computer and Communication Engineering, University of Science and Technology Beijing, Beijing, China}

\begin{abstract}
ECG–text pretraining has emerged as a promising paradigm for transferable ECG representation learning and zero-shot diagnosis.
However, clinical reports usually provide only sparse phrase-level supervision, whereas ECG signals contain rich and redundant diagnostic evidence distributed across leads and time. 
Under such asymmetric supervision, cross-modal alignment can easily degenerate into coarse prototype matching, where a few dominant ECG tokens become shortcuts for matching diagnostic phrases while complementary evidence remains underutilized. 
Although existing approaches mitigate shortcut reliance by enriching supervision or perturbing ECG evidence, they do not consider how ECG–text correspondence should evolve when available evidence changes.
In this work, we propose NESA, a Non-stationary Evidence-aware Semantic Alignment framework that abandons static prototype matching in favor of an evolving alignment trajectory. 
Specifically, NESA first quantifies the alignment contribution of each ECG token by measuring its compatibility with the current cross-modal alignment state. It then progressively suppresses alignment-dominant ECG evidence by erasing the ECG tokens with the highest alignment contribution in the latent space.
To maintain stable ECG–text correspondence during progressive evidence removal, NESA constructs an evolving evidence-conditioned alignment trajectory by integrating text-derived clinical semantics, full-view ECG context, and residual ECG evidence, providing adaptive alignment states and shaping the ECG encoder to learn representations that preserve semantic correspondence.
Experiments on MIMIC-IV-ECG pretraining and three downstream benchmarks demonstrate that NESA improves zero-shot diagnosis and label-efficient transfer. Dominant-token deletion analysis further shows that NESA maintains stable ECG--text alignment after removing highly aligned ECG evidence, supporting distributed ECG--text grounding.
\end{abstract}
\end{frontmatter}

\begin{figure}
\centering
\includegraphics[width=\textwidth]{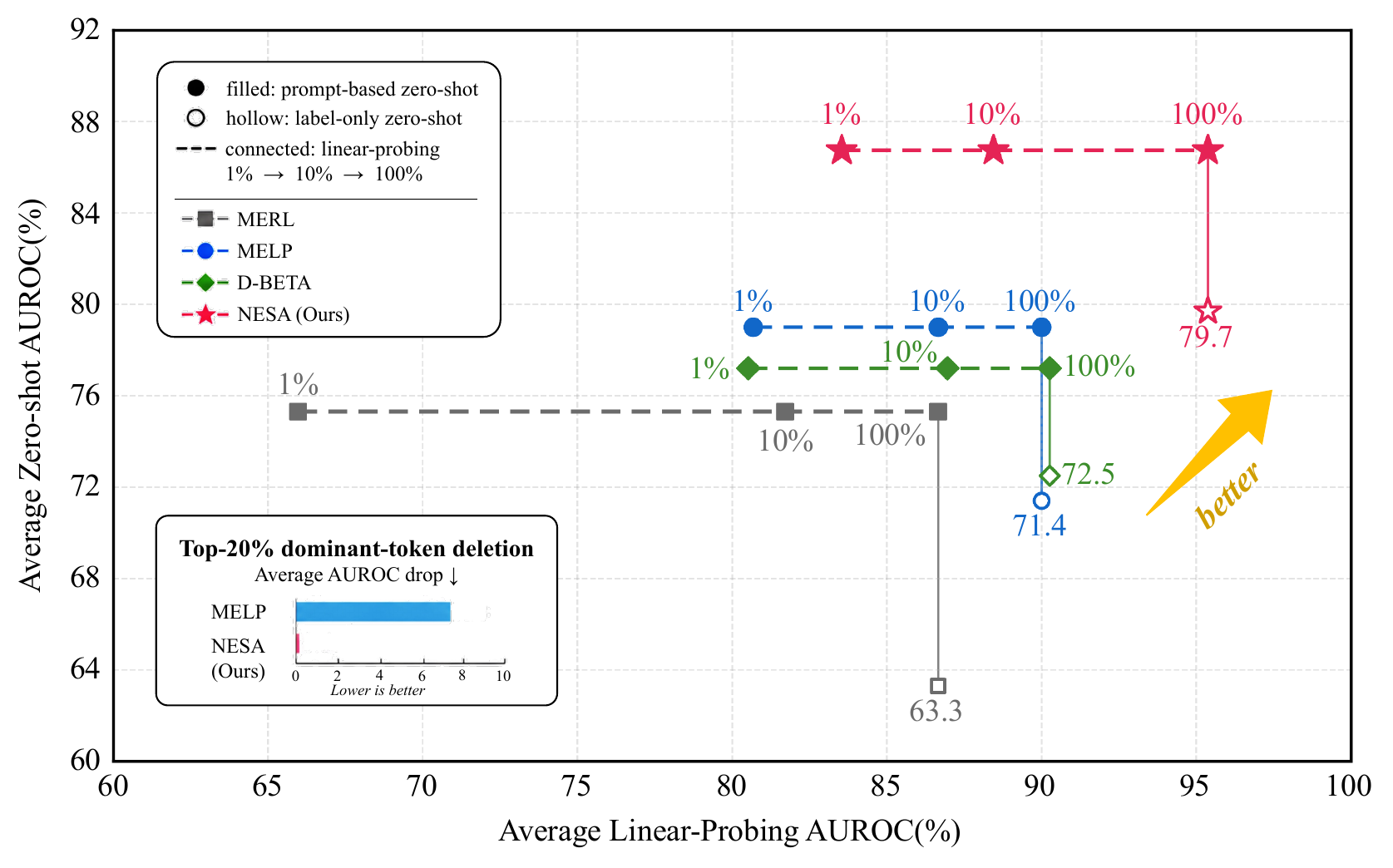}
% \caption{Comparison of zero-shot alignment, transferable representation quality, and robustness to sparse-text supervision. Each plotted method has results for both zero-shot AUROC and linear-probing AUROC, averaged across six evaluation tasks from three downstream ECG datasets. Connected points indicate the 1\%, 10\%, and 100\% linear-probing label fractions; hollow markers indicate label-only zero-shot performance. The inset reports performance under top-20\% dominant-token deletion. All competing methods shown are independent baselines evaluated under the stated protocols.}
\caption{
Comparison of zero-shot alignment, transferable representation quality,
and robustness to sparse-text supervision.
Each plotted method has results for both zero-shot AUROC and linear-probing AUROC,
averaged across six evaluation tasks from three downstream ECG datasets.
Connected points indicate the 1\%, 10\%, and 100\% linear-probing label fractions;
hollow markers indicate label-only zero-shot performance.
The inset reports performance under top-20\% dominant-token deletion
for methods with compatible token-level attribution protocols.
}
\label{fig:comparison}
\end{figure}

\begin{figure}
\centering
\includegraphics[width=\textwidth]{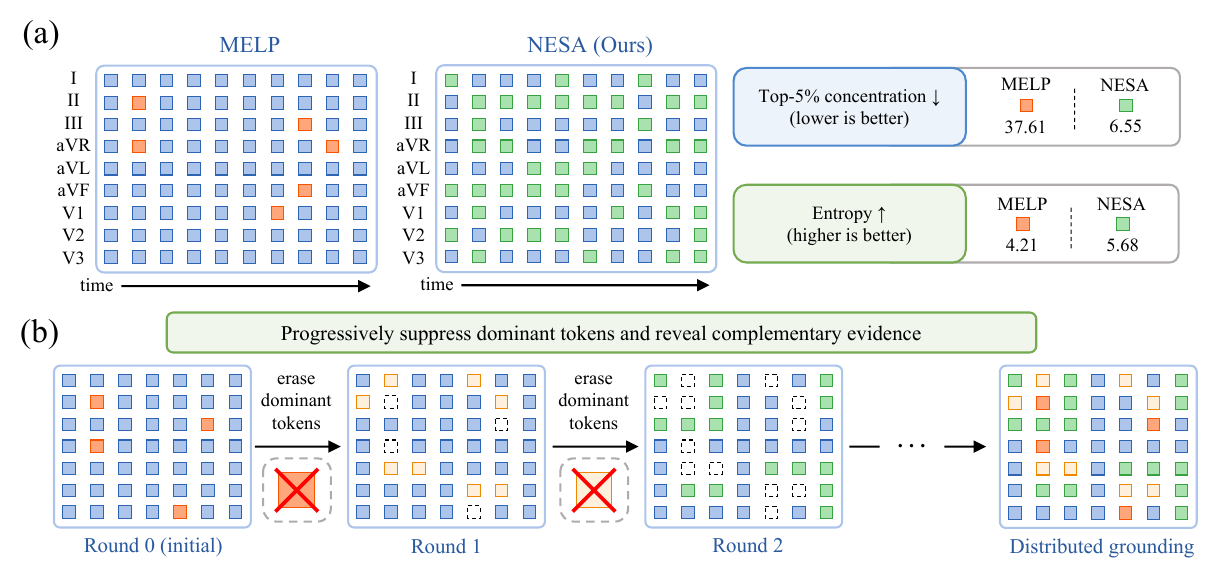}
% \caption{Motivation and empirical evidence of distributed ECG grounding under sparse supervision. 
% (a) ECG signals contain rich and redundant diagnostic evidence distributed across leads and time, while paired clinical text provides only sparse phrase-level supervision, leading to a strong supervision imbalance. 
% (b) We compare representative ECG--text pretraining methods. Existing approaches such as MELP and SCAR tend to concentrate alignment on a small subset of highly salient tokens, resulting in high evidence concentration and limited coverage. In contrast, NESA (ours) encourages more uniform utilization of ECG tokens, leading to lower top-k concentration and higher entropy of attention distribution. 
% (c) NESA progressively suppresses dominant tokens across multiple rounds of alignment refinement, forcing the model to discover complementary diagnostic evidence from previously underutilized regions, and ultimately yielding distributed ECG--text grounding.}
\caption{
Motivation and empirical evidence of distributed ECG grounding under sparse report supervision.
(a) Token-importance distributions of MELP and NESA measured with the same attribution procedure.
The corresponding top-5\% concentration and entropy summarize evidence coverage.
(b) NESA progressively suppresses dominant tokens across multiple rounds and uncovers complementary evidence from remaining ECG regions.
}
\label{fig:motivation}
\end{figure}

\section{Introduction}\label{sec:introduction}

Electrocardiogram (ECG) analysis plays a critical role in cardiovascular disease screening and diagnosis. Conventional deep learning approaches have achieved promising performance in various ECG-related tasks, but their generalization ability remains limited by dependence on task-specific annotations and predefined label spaces \citep{mckeen2024ecgfm}.
To overcome these limitations, recent studies have explored large-scale ECG representation learning, aiming to learn transferable representations that can adapt to diverse downstream diagnostic scenarios.
Among these approaches, ECG–text pretraining has emerged as a promising multimodal learning paradigm by leveraging clinical reports as scalable supervision for ECG representation learning \citep{ansari2025survey}.
However, clinical reports and ECG signals provide inherently asymmetric supervision: clinical reports typically consist of short and semi-structured diagnostic conclusions without detailed descriptions of waveform-level evidence, whereas ECG recordings contain rich and redundant diagnostic information distributed across multiple leads and temporal regions \citep{tang2025electrocardiogram}. 
This supervision asymmetry creates a fundamental challenge for ECG--text alignment (Fig.~\ref{fig:motivation}a), as sparse textual descriptions may not fully capture the diverse diagnostic evidence encoded in ECG signals.

Existing efforts to address this supervision mismatch have mainly followed two directions. One line of work enriches textual supervision by introducing external clinical descriptions or class semantics through prompting-based strategies. Another line of work perturbs ECG observations through masking or evidence suppression strategies to encourage broader utilization of ECG evidence.
However, these approaches still formulate ECG–text alignment as a static matching problem, overlooking the need for adaptive alignment when the underlying ECG evidence distribution changes
Consequently, the pretraining model may rely on a limited subset of highly aligned ECG evidence.

To break this shortcut reliance, an intuitive strategy is to explicitly suppress dominant ECG evidence, thereby forcing the model to discover complementary cues.
However, removing dominant evidence introduces a new challenge that has been largely overlooked: the available evidence distribution evolves during the cross-modal alignment. 
As the most informative ECG tokens are progressively removed, the remaining signal provides a different evidence context for supporting the same diagnostic concept. 
Therefore, continuously enforcing alignment with a fixed ECG-to-semantic correspondence may cause semantic inconsistency, since the optimal alignment state should adapt to the evolving evidence distribution.

Motivated by this observation, we propose NESA, a Non-stationary Evidence-aware Semantic Alignment framework that transforms ECG-text pretraining from static alignment optimization into an evolving alignment process. To actively break shortcut reliance, NESA progressively suppresses alignment-dominant ECG evidence in the latent space, encouraging the model to uncover complementary diagnostic cues from previously underutilized regions. Crucially, to address the challenge of shifting evidence availability caused by progressive removal, NESA introduces a gated alignment corrector that maintains an evolving evidence-conditioned alignment trajectory. Instead of forcing residual ECG representations to follow a fixed alignment state derived from complete evidence, the corrector progressively refines the alignment state at each round according to the currently available ECG evidence while preserving the original clinical semantics.

The main contributions of this work are summarized as follows:
\begin{itemize}
    \item We identify the supervision mismatch in ECG-text pretraining that causes shortcut reliance, and rethink cross-modal representation learning from the perspective of alignment stability under non-stationary evidence availability.
    \item We propose NESA, which progressively suppresses alignment-dominant ECG tokens through alignment-aware evidence scoring and introduces an adaptive alignment correction mechanism that dynamically updates the cross-modal anchor under evolving evidence states.
    \item Extensive experiments demonstrate NESA's superiority in zero-shot diagnosis and label-efficient transfer, with dominant-token deletion analysis verifying its distributed and robust ECG grounding. 
\end{itemize}

\section{Related Work} \label{sec:related_work}

\subsection{ECG Representation Learning and ECG--Text Pretraining}

Deep learning has been widely applied to ECG analysis, including arrhythmia classification, risk prediction, and foundation-level representation learning. Early ECG representation learning methods mainly relied on supervised learning from expert-labeled datasets, where the goal was to directly learn discriminative features for downstream diagnosis \citep{hannun2019cardiologist,rajpurkar2017cardiologist,acharya2017deep,fawaz2020inceptiontime}. To reduce dependence on labels and improve transferability, later studies introduced self-supervised objectives\citep{darban2025carla}. For example, CLOCS learns ECG representations through contrastive learning across space, time, and patients \citep{kiyasseh2021clocs}, while ST-MEM uses masked representation learning to capture spatio-temporal structure in ECG signals \citep{na2024stmem}. Other recent ECG foundation models further show that large-scale pretraining can substantially improve transfer performance across downstream tasks \citep{mckeen2024ecgfm}.

More recently, ECG--text pretraining has emerged as a promising direction by aligning ECG recordings with paired reports, diagnostic statements, or class prompts in a shared embedding space \citep{mckeen2024ecgfm,liu2024merl,wang2025melp,tang2026interpretable,radford2021clip}. To compensate for the sparsity of textual supervision, prompting-based methods attempt to enrich the text side by introducing external clinical descriptions. For instance, MERL utilizes clinical knowledge-enhanced prompting for zero-shot classification \citep{liu2024merl}, and MELP further improves this paradigm via multi-scale ECG-language alignment \citep{wang2025melp}. While these methods demonstrate that enriched textual supervision can serve as a clinically meaningful anchor, they are developed under the implicit assumption of a static alignment objective. In practice, ECG reports are often short and semi-structured, mainly consisting of phrase-level diagnostic conclusions without detailed waveform-level evidence \citep{zhang2026meeti,gow2023mimic,zhao2025ecg}. Under this supervision imbalance, optimizing toward a fixed textual target leaves models vulnerable to shortcut learning.

Beyond ECG-specific models, cross-modal learning methods have examined how feature granularity and modality-specific redundancy affect semantic correspondence. Cross-modal alignment learning for text-based person search suppresses redundant modality-specific information while aligning image and text features at multiple granularities \citep{KE2024110481}. CSA aligns semantic units across scales and filters weak associations \citep{LIU2025111647}. In medical image--text learning, self-adaptive image--text fusion selects strongly correlated visual and report features \citep{ZHANG2025111715}. Clinically informed prompt learning anchors prompt tokens to diagnostic attributes \citep{XIE2026114114}, while morphology semantics-guided alignment links visual evidence to class-level descriptions \citep{LEI2026114606}. DYC-CLIP uses local visual context to adapt prompts for zero-shot anomaly detection \citep{CHEN2026113215}. Building on these studies, NESA examines how ECG--text alignment should evolve as dominant ECG evidence is progressively removed.

\subsection{Mask-Based Pretraining and Learning from Incomplete Observations}

Mask-based learning has become a core paradigm in representation learning across vision\citep{xing2026ma}, language\citep{yang2024continual}, and time-series domains\citep{li2024learning}. In ECG and other time-series applications, masking and reconstruction are commonly used to learn local and global structure from partially observed inputs. ST-MEM, for example, explicitly guides masked representation learning to capture spatio-temporal ECG relationships \citep{na2024stmem}, while other self-supervised ECG methods also rely on partial observation to improve label efficiency and transferability \citep{kiyasseh2021clocs}. Similar trends can be found in multimodal learning, where masked objectives and modality dropout are often used to improve robustness under missing or corrupted inputs.

Although these methods show that learning from incomplete observations can force broader signal utilization and produce more transferable representations, most existing masking strategies rely on random dropping and remain agnostic to the semantic role of the removed content \citep{quan2025semantic}. In computer vision, researchers \citep{wang2023hpm} argue that not all masked regions contribute equally to representation learning, and that harder or more informative patches provide stronger training signals than random masking. A similar limitation arises in ECG--text pretraining. When the paired text is sparse, random masking often leaves the easiest alignment path intact, allowing the model to continue relying on a small subset of highly salient ECG tokens. As a result, much of the remaining ECG evidence is still weakly utilized. This motivates going beyond generic random masking and instead explicitly targeting the ECG evidence that currently dominates cross-modal alignment.

% In the ECG domain, recent work such as SCAR explores adversarial masking of alignment-critical ECG regions to improve robustness under partial observations \citep{liu2026scar}. These methods primarily focus on identifying and removing salient evidence to enhance robustness, but typically treat the alignment objective as fixed during the masking process. However, in ECG--text pretraining under sparse phrase-level supervision, the removal of dominant evidence may also alter the effective alignment context, suggesting the need to consider how cross-modal alignment should evolve under progressively changing evidence availability.
\subsection{Adversarial Erasing, Counterfactual Intervention, and Shortcut Reduction}

A related line of research explores adversarial erasing and counterfactual intervention, aiming to reduce reliance on highly discriminative but potentially spurious features. In weakly supervised localization and segmentation, adversarial erasing methods remove the most salient image regions to encourage the discovery of complementary evidence \citep{wei2017ae}. Similar ideas have been extended to multimodal reasoning tasks, where counterfactual intervention is used to construct perturbed samples for robustness improvement. For example, Counterfactual Samples Synthesizing masks critical objects to improve robustness in VQA \citep{chen2020css}, and Efficient Counterfactual Debiasing applies selective masking of objects or words to mitigate shortcut dependence \citep{kolling2022ecvqa}. These approaches demonstrate that suppressing highly influential evidence can encourage models to utilize broader contextual information.

In the ECG domain, recent evidence suppression methods explore adversarial
masking of alignment-critical ECG regions to improve robustness under partial
observations. These methods primarily focus on identifying and removing
salient evidence to enhance robustness, but typically treat the alignment
objective as fixed during the masking process. However, in ECG--text
pretraining under sparse phrase-level supervision, the removal of dominant
evidence may also alter the effective alignment context, suggesting the need
to consider how cross-modal alignment should evolve under progressively
changing evidence availability.

\begin{figure}
\centering
\includegraphics[width=\textwidth]{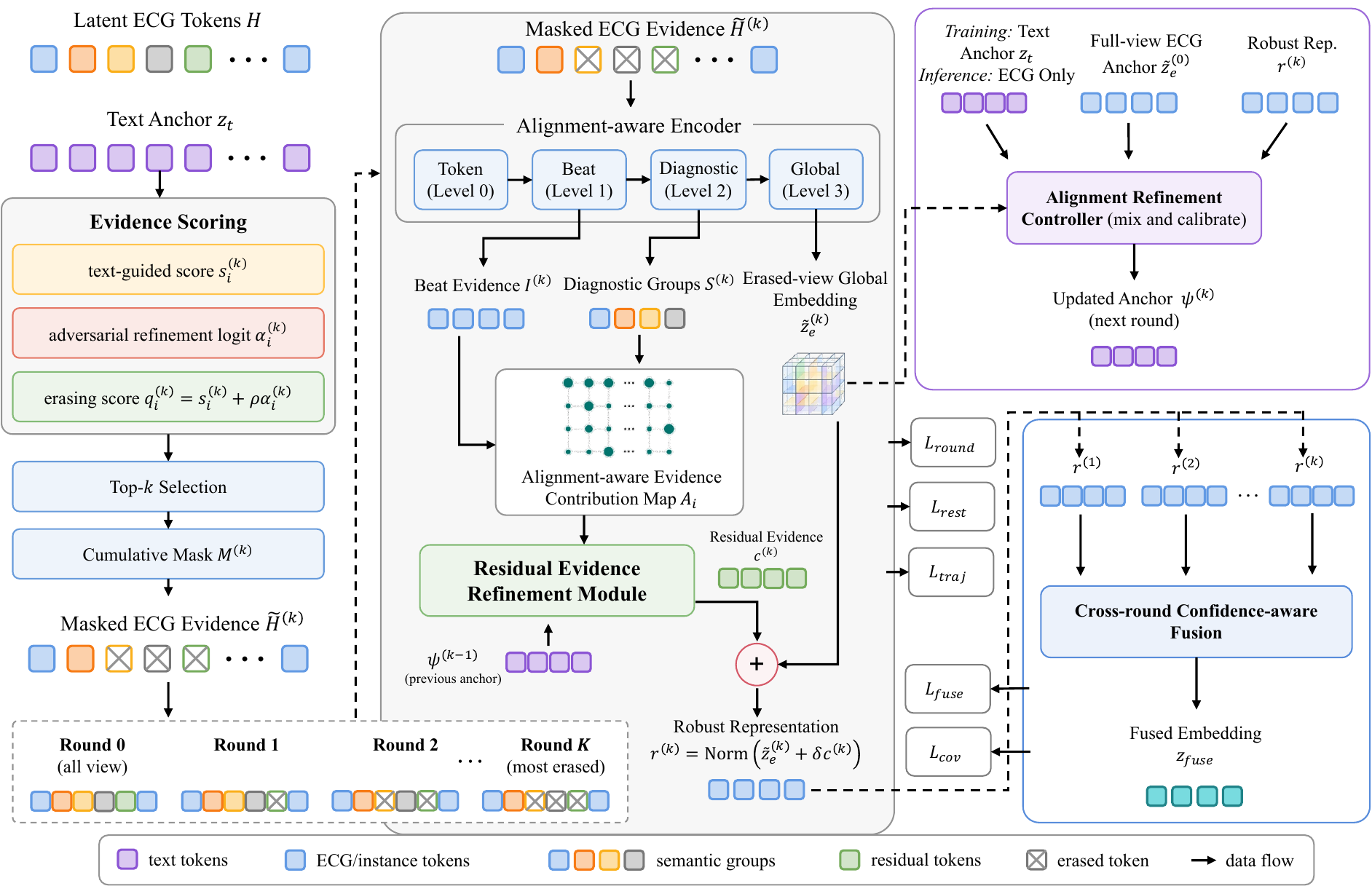}
\caption{Overview of NESA. NESA progressively suppresses alignment-dominant latent ECG tokens and re-aggregates the remaining evidence through an alignment-aware encoder. The encoder organizes suppressed ECG evidence into beat-level evidence, diagnostic groups, and a global alignment embedding. A residual evidence refinement module recovers complementary cues under the previous anchor, producing robust round representations. The alignment refinement controller updates the anchor for the next round, and iterative alignment fusion integrates all rounds into the final ECG embedding.}
\label{fig:model_architecture}
\end{figure}

\section{Method}

\subsection{Task Formulation}
\label{sec:problem_definition}

Let $\mathcal{D}=\{(x_i,t_i)\}_{i=1}^{N}$ denote a paired ECG--text pretraining dataset, where $x_i\in\mathbb{R}^{C\times T}$ is a $C$-lead ECG recording and $t_i$ is the paired clinical report, diagnostic statement, or class prompt. 
ECG--text pretraining learns an ECG encoder and a text encoder that align paired ECG--text samples in a shared semantic space, enabling semantic correspondence between cardiac signals and clinical concepts.

To enable token-level evidence modeling, we employ an ECG backbone to extract latent representations from the multi-lead ECG signal.
Specifically, the ECG backbone maps the input $x\in\mathbb{R}^{C\times T}$ into a sequence of $M$ contextualized representations:
\begin{equation}
    H_b=f_{\mathrm{ECGFM}}(x) \in\mathbb{R}^{M\times d_b},
\end{equation}
where $M$ denotes the number of backbone tokens and $d_b$ is the backbone feature dimension.
These representations are further projected into a shared latent space, yielding $\mathcal{H}=\{h_i\}_{i=1}^{M}$, where $h_i\in\mathbb{R}^{d}$.
Each latent token captures contextual information from the multi-lead ECG signal, providing a representation-level basis for estimating alignment contributions and performing
progressive evidence suppression.

The ECG encoder is written as a tokenization backbone $f_e^{\mathrm{tok}}(\cdot)$ followed by a shared aggregation encoder $f_e^{\mathrm{agg}}(\cdot)$. The full-view ECG embedding is $z_e^{(0)}=f_e^{\mathrm{agg}}(\mathcal{H})$, and the paired text embedding is $z_t=f_t(t)$. The standard full-view ECG--text alignment objective is
\begin{equation}
\mathcal{L}_{\mathrm{align}}
=
-\log
\frac{\exp(\mathrm{sim}(z_e^{(0)},z_t)/\tau_a)}
{\sum_{t'\in\mathcal{B}}\exp(\mathrm{sim}(z_e^{(0)},z_{t'})/\tau_a)},
\label{eq:align}
\end{equation}
where $\mathrm{sim}(\cdot,\cdot)$ denotes cosine similarity, $\tau_a$ is the contrastive temperature, and $\mathcal{B}$ denotes the mini-batch. When report reconstruction is enabled, an auxiliary caption-generation loss $\mathcal{L}_{\mathrm{cap}}$ is also used.

As illustrated in Fig.~\ref{fig:model_architecture}, NESA performs iterative evidence-aware alignment over contextualized ECG tokens. Given the full-view ECG representation, NESA first estimates token-level alignment contributions and progressively suppresses dominant evidence to generate different evidence states. The masked representations are then encoded to recover complementary diagnostic information, while a gated alignment correction module updates the alignment anchor according to the changed evidence availability. During pretraining, the initial anchor is initialized from the paired report embedding, whereas during zero-shot inference it is initialized from the full-view ECG representation due to the absence of paired text or labels. In both cases, the anchor is iteratively refined with stop-gradient applied to the input representations, and representations from different rounds are fused to obtain the final robust ECG embedding.

\subsection{Alignment-Aware Evidence Scoring}
\label{sec:evidence_scoring}

After evidence removal, the available ECG information changes progressively, and the tokens that dominate the ECG--text alignment may also shift across rounds. 
Therefore, instead of assigning a fixed importance score to each ECG token, NESA estimates token contributions conditioned on the current alignment state. 
Specifically, we maintain a dynamic alignment anchor $\psi^{(k-1)}$ at each erasing round $k$, which summarizes the current semantic state after previous evidence suppression. 
Given the remaining token set $\Omega^{(k)}$, we summarize the available ECG evidence by its mean representation, $\bar{h}_{\mathrm{res}}^{(k)}=|\Omega^{(k)}|^{-1}\sum_{i\in\Omega^{(k)}}h_i$.

Conditioned on the current alignment state, each remaining token is evaluated from two complementary perspectives. 
The first component measures how strongly a token supports the current cross-modal alignment anchor $\psi^{(k-1)}$, while the second identifies removals that create a challenging counterfactual condition. They are computed as $s_i^{(k)}=g_{\mathrm{evd}}(h_i,\psi^{(k-1)},\bar{h}_{\mathrm{res}}^{(k)})$ and $a_i^{(k)}=g_{\mathrm{adv}}(h_i,\psi^{(k-1)},\bar{h}_{\mathrm{res}}^{(k)})$ for $i\in\Omega^{(k)}$.

Here, $g_{\mathrm{evd}}$ and $g_{\mathrm{adv}}$ are lightweight MLP-based scoring heads. 
The evidence scorer $g_{\mathrm{evd}}$ estimates the alignment contribution of each token under the current semantic state. 
The adversarial refiner $g_{\mathrm{adv}}$ provides an additional refinement signal by learning which token removals can most effectively perturb the current alignment anchor.
The final evidence suppression score is defined as
\begin{equation}
q_i^{(k)}
=
s_i^{(k)}
+
\rho a_i^{(k)},
\qquad
i\in\Omega^{(k)},
\label{eq:final_erasing_score}
\end{equation}
where $\rho$ controls the strength of adversarial refinement. 
Tokens with higher $q_i^{(k)}$ are considered dominant under the current alignment condition and are selected for suppression in the following stage.
For cross-round coverage regularization, the evidence scores are normalized as $w_i^{(k)}=\mathrm{softmax}_{i\in\Omega^{(k)}}(s_i^{(k)}/\tau_e)$.

The adversarial refiner is optimized with an alternating training strategy. 
During the refiner update, the ECG and text encoders are detached, and the refiner learns to identify counterfactual token selections whose removal maximally challenges the current alignment state. 
During the main model update, the selected suppressed views are treated as hard evidence perturbations, encouraging the ECG--text model to recover semantically consistent representations from incomplete evidence.

\subsection{Progressive Evidence Suppression and Alignment-Aware Encoding}
\label{sec:progressive_erasing}

Based on the alignment-aware evidence scores defined in Sec.~\ref{sec:evidence_scoring}, NESA progressively suppresses ECG evidence that dominates the current cross-modal alignment state. 
Different from fixed masking strategies that perturb ECG representations without considering their semantic roles, NESA performs state-dependent counterfactual erasing, where the removed evidence is determined according to the evolving alignment anchor.

Formally, let $\mathcal{H}=\{h_i\}_{i=1}^{M}$ denote the complete latent ECG token set and $\mathcal{M}^{(k)}\in\{0,1\}^{M}$ the cumulative mask at round $k$. Token-wise masking gives the remaining evidence $\tilde{\mathcal{H}}^{(k)}=(1-\mathcal{M}^{(k)})\odot\mathcal{H}$.
The suppression trajectory is recursively updated as
\begin{equation}
\mathcal{M}^{(k)}
=
\mathcal{M}^{(k-1)}
\vee
\mathrm{TopK}
\left(
q^{(k)}
(\mathcal{H}\setminus\mathcal{M}^{(k-1)})
\right),
\label{eq:mask_update}
\end{equation}
where $q^{(k)}$ denotes the alignment-aware erasing score defined in Sec.~\ref{sec:evidence_scoring}. 
The $\mathrm{TopK}(\cdot)$ operation selects the currently dominant ECG tokens according to the current alignment state. 
Since the scoring function is conditioned on $\psi^{(k-1)}$, each suppression round removes evidence that is dominant under the current semantic condition.

To characterize the remaining evidence after counterfactual suppression, NESA introduces an alignment-aware ECG encoder operating on $\tilde{\mathcal{H}}^{(k)}$:
\begin{equation}
\tilde{z}_e^{(k)}
=
f_{\mathrm{enc}}
\left(
\tilde{\mathcal{H}}^{(k)},
\psi^{(k-1)}
\right),
\label{eq:alignment_encoding}
\end{equation}
where $f_{\mathrm{enc}}(\cdot)$ performs hierarchical evidence aggregation over contextualized ECG representations.
Specifically, the encoder first processes token-level representations, then aggregates related features into higher-level semantic representations, and finally produces an erased-view ECG representation conditioned on the current alignment state.

The resulting representations
$\{\tilde{z}_e^{(1)},\tilde{z}_e^{(2)},\ldots,\tilde{z}_e^{(K)}\}$
correspond to different evidence availability states. 
These counterfactual states expose how the semantic correspondence changes when dominant evidence is progressively removed, providing the basis for the subsequent gated alignment correction.

\subsection{Gated Alignment Correction and Fusion}
\label{sec:comp_restoration}

Progressive evidence suppression generates a sequence of ECG representations with different evidence availability. 
However, removing alignment-dominant evidence also changes the semantic context available for cross-modal matching, making a fixed alignment pathway insufficient. 
To address this issue, NESA introduces a gated alignment correction module that dynamically updates the alignment anchor according to the evolving evidence distribution.

Given the erased-view representation $\tilde{z}_e^{(k)}$, NESA aggregates complementary residual evidence as $c^{(k)}=\Delta(\tilde{\mathcal{H}}^{(k)},\psi^{(k-1)})$, where $\Delta(\cdot)$ emphasizes evidence that becomes informative after dominant-token suppression.
The round-wise robust ECG representation is then obtained by integrating the erased-view representation and residual evidence:
\begin{equation}
r^{(k)}
=
\mathrm{Norm}
\left(
\tilde{z}_e^{(k)}
+
\delta c^{(k)}
\right),
\label{eq:robust_rep}
\end{equation}
where $\delta$ controls the contribution of residual evidence refinement.

Instead of forcing the updated ECG representation to match a fixed textual target, the gated alignment correction module dynamically updates the cross-modal anchor by combining three complementary sources: the original text semantics, the full-view ECG representation, and the current refined representation.

During training, the gate weights are $\boldsymbol{\mu}_{\mathrm{train}}^{(k)}=\mathrm{Softmax}(\mathrm{MLP}_{\mathrm{train}}([r^{(k)},z_t,z_e^{(0)}]))$, and the alignment anchor is updated as
\begin{equation}
\psi_{\mathrm{train}}^{(k)}
=
\mathrm{Norm}
\left(
\mu_t^{(k)}\mathrm{sg}(z_t)
+
\mu_e^{(k)}\mathrm{sg}(z_e^{(0)})
+
\mu_r^{(k)}\mathrm{sg}(r^{(k)})
\right).
\end{equation}
Here, $\mathrm{sg}(\cdot)$ denotes stop-gradient.
During zero-shot inference, NESA uses only ECG-side representations. The gate weights are $\boldsymbol{\mu}_{\mathrm{test}}^{(k)}=\mathrm{Softmax}(\mathrm{MLP}_{\mathrm{test}}([r^{(k)},z_e^{(0)}]))$, and the anchor becomes $\psi_{\mathrm{test}}^{(k)}=\mathrm{Norm}(\mu_e^{(k)}\mathrm{sg}(z_e^{(0)})+\mu_r^{(k)}\mathrm{sg}(r^{(k)}))$.

% To maintain semantic continuity along the evolving alignment trajectory, we introduce a cross-round consistency constraint:

To maintain semantic continuity along the evolving alignment trajectory, we introduce a cross-round semantic consistency loss:
\begin{equation}
\mathcal{L}_{\mathrm{traj}}
=
\frac{1}{K}
\sum_{k=1}^{K}
\left[
1-
\mathrm{sim}
(
r^{(k)},
\mathrm{sg}(\psi_{\mathrm{train}}^{(k-1)})
)
\right].
\label{eq:trajectory}
\end{equation}

Finally, NESA aggregates round-wise representations through confidence-aware fusion as $z_{\mathrm{fuse}}=\sum_{k=1}^{K}\pi_k r^{(k)}$, where $\pi_k=\mathrm{Softmax}(g_{\mathrm{fuse}}(r^{(k)},\psi^{(k)}))$.
The fused representation integrates complementary evidence discovered under different alignment states, enabling robust ECG--text alignment under non-stationary evidence availability.
For zero-shot classification, we compute the cosine similarity between $z_{\mathrm{fuse}}$ and the fixed class-prompt embeddings $\{z_{c_1},z_{c_2},\ldots,z_{c_K}\}$ to obtain diagnostic logits.

\section{Experiments}\label{sec:experiments}

\subsection{Experimental Setup}
\label{sec:exp_setup}

% \paragraph{Pretraining dataset.}
% Following recent ECG--text pretraining studies, all methods are pretrained on MIMIC-IV-ECG, a large-scale paired ECG--text corpus containing 12-lead clinical ECG recordings and corresponding diagnostic reports. The pretraining set is used only for multimodal representation learning, while all downstream evaluations are conducted on public benchmarks without using their test labels during pretraining.
\paragraph{Pretraining dataset.}
NESA is pretrained on MIMIC-IV-ECG, a large-scale paired ECG--text corpus
containing 12-lead clinical ECG recordings and corresponding diagnostic
reports. Baseline results follow the training settings and model sources of
their respective implementations.

\begin{table}[t]
\centering
\caption{Summary of datasets used for pretraining and downstream evaluation.}
\label{tab:dataset_summary}
\small
\begin{tabular}{lccccc}
\toprule
Dataset & Usage & ECGs & Subjects & Format & Labels/Tasks\\
\midrule
MIMIC-IV-ECG & Pretraining & 800,035 & 161,352 & 12-lead, 500Hz & ECG-report pairs\\
PTB-XL & Downstream & 21,837 & 18,885 & 12-lead, 500Hz & 5/23/19/12 labels\\
CPSC2018 & Downstream & 6,877 & 6,877 & 12-lead, 500Hz & 9 labels\\
CSN & Downstream & 23,026 & 23,026 & 12-lead, 500Hz & 38 labels\\
\bottomrule
\end{tabular}
\end{table}

\paragraph{Downstream datasets.}
We evaluate the pretrained models on three widely used ECG benchmarks: PTB-XL, CPSC2018, and Chapman--Shaoxing--Ningbo (CSN). PTB-XL is further evaluated on its standard multi-label subsets, including Superclass, Subclass, Form, and Rhythm. Table~\ref{tab:dataset_summary} summarizes the main statistics of all datasets used in this study. Detailed dataset preprocessing and split strategies are provided in Appendix.

\paragraph{Evaluation protocols.}
We consider two standard downstream protocols. 
(1) \textbf{Zero-shot classification} matches ECG embeddings with the fixed clinically enriched prompts (CKEPE) released by MERL.
(2) \textbf{Linear probing} evaluates the transferability of the pretrained ECG representation by freezing the ECG encoder and training a linear classifier on top of the ECG embedding. 
To further examine the proposed mechanism, we additionally conduct dominant-token deletion analysis, evidence coverage analysis, qualitative token-importance visualization, ablation studies, and hyperparameter sensitivity experiments.

\paragraph{Label-only sparse-text protocol.}
In addition to the standard zero-shot setting, we further evaluate a stricter \textbf{label-only} protocol, in which each class is represented only by its short diagnostic phrase or label name without prompt enrichment. This setting provides an explicit test of robustness under extremely sparse textual supervision.

% \paragraph{Baselines.}
% We compare NESA with representative ECG--text pretraining methods, including MERL, MELP, and SCAR, in zero-shot classification. For linear probing, we further include representative ECG pretraining baselines such as CLOCS, ST-MEM, ECGFM, and HeartLang. For all ECG--text baselines, we use the same downstream splits and identical prompt inputs under each evaluation protocol. For linear probing, all compared methods use frozen ECG representations and the same classifier training protocol.
\paragraph{Baselines.}
For zero-shot classification, we compare NESA with ECG--text
pretraining methods, including MERL, MELP, D-BETA, and ZETA. For linear
probing, we additionally evaluate representative ECG representation-learning
methods, including CLOCS, ST-MEM, ECGFM, and HeartLang, together with ECG--text
pretraining approaches when applicable.

% \paragraph{Implementation details.}
% NESA adopts ECGFM-small as the ECG backbone and 
% \texttt{ncbi/MedCPT-Query-Encoder} as the text encoder.
% The ECG signal is partitioned into lead--time tokens by dividing each lead into temporal patches.
% The backbone tokens are projected into a 256-dimensional latent space, where progressive evidence erasing, alignment-aware scoring, and token-level evidence analysis are performed.
% The multimodal model is pretrained for up to 100 epochs with mixed precision, while the first six layers of the text encoder are frozen. Unless otherwise specified, we use $K=3$ progressive erasing rounds and a ratio-based erasing budget of $r=5\%$ per round. Progressive evidence erasing is only used during training. 
% During zero-shot inference, all class text embeddings are computed once from the predefined prompt set and kept fixed. NESA performs sample-wise evidence refinement using only the input ECG and fixed model parameters, without accessing ground-truth labels, label vectors, test-set statistics, or performing test-time adaptation.
\paragraph{Implementation details.}
NESA adopts ECGFM-small as the ECG backbone and 
\texttt{ncbi/MedCPT-Query-Encoder} as the text encoder.
The ECG backbone extracts contextualized token representations from the multi-lead ECG signal through convolutional feature extraction and Transformer encoding.
The backbone tokens are projected into a 256-dimensional latent space, where progressive evidence suppression, alignment-aware scoring, and token-level evidence analysis are performed.
The multimodal model is pretrained for up to 100 epochs with mixed precision, while the first six layers of the text encoder are frozen. Unless otherwise specified, we use $K=3$ progressive suppression rounds and a ratio-based suppression budget of $r=5\%$ per round.

Progressive evidence suppression is applied during both pretraining and zero-shot inference. During pretraining, multiple evidence states generated through progressive suppression are used to optimize the non-stationary alignment objective. During zero-shot inference, NESA performs sample-wise evidence refinement by iteratively suppressing dominant ECG tokens and updating the ECG--text alignment trajectory with fixed model parameters.

% During inference, NESA supports two evaluation paths. For linear probing and encoder-only transfer, the progressive erasing, restoration, and fusion branches are discarded, and only the pretrained ECG encoder is retained. For zero-shot classification, all class text embeddings are computed once from the predefined prompt set and kept fixed throughout evaluation. Given a test ECG, NESA first computes its full-view ECG embedding and initializes the erasing anchor from the ECG side only. The subsequent multi-round erasing, compensatory restoration, alignment correction, and round fusion are performed independently for each test sample, using only its ECG representation and the fixed model parameters. No ground-truth label, label vector, test-set statistic, batch-level information, or parameter update is used during this process. The final fused ECG embedding is then matched with the same fixed class text embeddings to obtain zero-shot logits. Thus, the reported zero-shot results follow a sample-wise deterministic inference protocol.

\subsection{Main Results}\label{sec:main_results}

% \paragraph{Zero-shot classification.}
% Table~\ref{tab:main_zeroshot} reports standard zero-shot AUROC across multiple downstream datasets. NESA achieves the best average AUROC of 86.7\%, improving over the strongest baseline SCAR (82.9\%) by 3.8 points. Specifically, NESA obtains the best performance on CSN (82.6\%), PTBXL-Rhythm (89.2\%), PTBXL-Sub (85.4\%), PTBXL-Super (96.8\%), and CPSC2018 (94.4\%), and remains competitive on PTBXL-Form (72.1\%). These results show that NESA consistently improves cross-dataset zero-shot ECG--text alignment.
\paragraph{Zero-shot classification.}
Table~\ref{tab:main_zeroshot} reports standard zero-shot AUROC across
multiple downstream datasets. NESA achieves the best average AUROC of
86.7\%, improving over the strongest baseline ZETA (79.2\%)
by 7.5 percentage points. Specifically, NESA obtains the best performance
on CSN (82.6\%), PTBXL-Rhythm (89.2\%), PTBXL-Sub (85.4\%),
PTBXL-Super (96.8\%), and CPSC2018 (94.4\%), while achieving competitive
performance on PTBXL-Form (72.1\%), where ZETA obtains the highest AUROC.
These results demonstrate consistent improvements across the downstream
evaluation tasks.
% \paragraph{Robustness under label-only supervision.}
% Table~\ref{tab:main_zeroshot_label} reports a stricter label-only zero-shot evaluation, where each class is represented only by its short diagnostic phrase without prompt enrichment. This setting provides substantially weaker textual supervision and therefore constitutes a more demanding test of ECG--text grounding. Under this protocol, NESA achieves the best overall average AUROC of 79.7\%, with an average drop of 7.0 points from the prompt-based setting. NESA obtains the best performance on CSN (69.3\%), PTBXL-Rhythm (90.1\%), PTBXL-Form (72.5\%), PTBXL-Sub (80.7\%), and PTBXL-Super (86.9\%), while SCAR performs better on CPSC2018. These results show that NESA remains robust when textual semantics are compressed to label-only supervision, supporting our claim that progressive evidence refinement reduces shortcut token--text matching and encourages more stable grounding in complementary ECG cues.
\paragraph{Robustness under label-only supervision.}
Table~\ref{tab:main_zeroshot_label} reports a stricter label-only
zero-shot evaluation, where each class is represented only by its
short diagnostic phrase without prompt enrichment. This setting
provides substantially less textual context, posing a more
challenging test of ECG--text alignment. Under this protocol,
NESA achieves the highest average AUROC of 79.7\%, outperforming
the strongest baseline result of 72.5\% (D-BETA and ZETA) by
7.2 percentage points. Specifically, NESA obtains the best
performance on five of the six evaluation tasks, while MELP
achieves the highest AUROC on CPSC2018 (79.8\% versus 78.6\%
for NESA). Compared with the prompt-based setting, NESA's
average AUROC decreases by 7.0 points. These results demonstrate
that NESA maintains strong zero-shot classification performance
even when textual descriptions are reduced to short diagnostic
phrases, suggesting robust ECG--text alignment under limited
semantic supervision. Because ZETA differs from D-BETA only in prompt construction, removing prompt enrichment makes the two methods identical under the label-only protocol, explaining their matching results in Table~4.

\begin{table*}
\centering
\caption{Linear probing performance (AUROC [\%]) of NESA and baseline models across multiple datasets. Results are reported for different training data proportions (1\%, 10\%, and 100\%). The best and second-best results are highlighted in bold and underlined, respectively.}
\label{tab:main_lp}
\resizebox{\textwidth}{!}{
\begin{tabular}{l|ccc|ccc|ccc|ccc|ccc|ccc}
\toprule
\multirow{2}{*}{Methods}
& \multicolumn{3}{c|}{PTBXL-Rhythm}
& \multicolumn{3}{c|}{PTBXL-Sub}
& \multicolumn{3}{c|}{PTBXL-Form}
& \multicolumn{3}{c|}{PTBXL-Super}
& \multicolumn{3}{c|}{CPSC2018}
& \multicolumn{3}{c}{CSN} \\
\cline{2-19}
& 1\% & 10\% & 100\%
& 1\% & 10\% & 100\%
& 1\% & 10\% & 100\%
& 1\% & 10\% & 100\%
& 1\% & 10\% & 100\%
& 1\% & 10\% & 100\% \\
\midrule
CLOCS
& 47.19 & 71.88 & 76.31
& 57.94 & 72.55 & 76.24
& 51.97 & 57.79 & 72.65
& 68.94 & 73.36 & 76.31
& 59.59 & 77.78 & 77.49
& 54.38 & 71.93 & 76.13 \\

ST-MEM
& 51.12 & 65.44 & 74.85
& 54.12 & 57.86 & 63.59
& 55.71 & 59.99 & 66.07
& 61.12 & 66.87 & 71.36
& 56.69 & 63.32 & 70.39
& 59.77 & 66.87 & 71.36 \\

ECGFM
& 81.45 & 91.59 & 92.70
& 73.24 & 81.91 & 86.07
& 60.95 & 74.99 & \underline{85.54}
& 78.67 & 84.80 & 86.47
& 82.18 & 89.52 & 93.26
& 71.51 & 83.17 & 88.89 \\

HeartLang
& 62.08 & 76.22 & 90.34
& 64.68 & 79.34 & \underline{88.91}
& 58.70 & 63.99 & 80.23
& 78.94 & 85.59 & 87.52
& 60.44 & 66.26 & 77.87
& 57.94 & 68.93 & 82.49 \\

MERL
& 53.33 & 82.88 & 88.34
& 64.90 & 80.59 & 84.72
& 58.26 & 72.43 & 79.65
& 82.39 & 86.27 & 88.67
& 70.33 & 85.32 & 90.57
& 66.60 & 82.74 & 87.95 \\

MELP
& \underline{88.83} & \underline{94.65} & \underline{96.91}
& \textbf{79.22} & \textbf{84.40} & 87.46
& 63.41 & 76.71 & 83.30
& 85.82 & 87.61 & 87.87
& \textbf{88.54} & \textbf{91.75} & 94.32
& \underline{78.25} & \underline{84.83} & 90.17 \\

D-BETA
& 83.15 & 88.36 & 90.11
& \underline{77.74} & 82.92 & 85.15
& \underline{70.10} & \textbf{78.91} & 83.98
& \underline{86.61} & \underline{92.83} & \underline{96.71}
& \underline{85.46} & \underline{91.35} & \underline{94.92}
& \textbf{80.04} & \textbf{87.36} & \underline{90.71} \\

\midrule
NESA (Ours)
& \textbf{94.77} & \textbf{98.52} & \textbf{99.37}
& 77.41 & \underline{84.33} & \textbf{93.27}
& \textbf{75.21} & \underline{78.04} & \textbf{87.44}
& \textbf{96.11} & \textbf{98.85} & \textbf{99.90}
& 85.03 & 89.34 & \textbf{98.33}
& 72.72 & 81.58 & \textbf{93.94} \\
\bottomrule
\end{tabular}
}
\end{table*}

% \paragraph{Linear probing.}
% Table~\ref{tab:main_lp} reports linear probing performance under 1\%, 10\%, and 100\% label fractions. NESA yields strong transfer performance, with particularly clear gains on several PTB-XL subsets and consistent best performance under full supervision. Under 1\% labels, NESA achieves the best results on PTBXL-Rhythm, PTBXL-Form, and PTBXL-Super, improving the strongest competing baselines from 88.83\% to 94.77\%, from 70.10\% to 75.21\%, and from 86.62\% to 96.11\%, respectively. Under full supervision, NESA also achieves the best results on all six evaluation subsets, including 99.37\% on PTBXL-Rhythm, 87.44\% on PTBXL-Form, 93.27\% on PTBXL-Sub, 99.90\% on PTBXL-Super, 98.33\% on CPSC2018, and 93.94\% on CSN. These results suggest that the benefit of progressive evidence erasing is not limited to zero-shot matching, but also improves the general transferability of the learned ECG representation.
\paragraph{Linear probing.}
Table~\ref{tab:main_lp} reports linear probing performance under 1\%, 10\%, and 100\% label fractions. NESA yields strong transfer performance, with particularly clear gains on several PTB-XL subsets and consistent best performance under full supervision. Under 1\% labels, NESA achieves the best results on PTBXL-Rhythm, PTBXL-Form, and PTBXL-Super, improving the strongest competing baselines from 88.83\% to 94.77\%, from 70.10\% to 75.21\%, and from 86.61\% to 96.11\%, respectively. Under full supervision, NESA also achieves the best results on all six evaluation tasks, including 99.37\% on PTBXL-Rhythm, 87.44\% on PTBXL-Form, 93.27\% on PTBXL-Sub, 99.90\% on PTBXL-Super, 98.33\% on CPSC2018, and 93.94\% on CSN. These results demonstrate that NESA learns transferable ECG representations that remain effective beyond zero-shot matching, achieving strong downstream performance across different levels of label availability.

\begin{table}
\centering
\caption{Zero-shot classification performance (AUROC [\%]) of NESA and baseline models across multiple datasets.}
\label{tab:main_zeroshot}
\resizebox{\columnwidth}{!}{
% \begin{tabular}{lccccccc}
% \toprule
% Methods & CSN & PTBXL-Rhythm & PTBXL-Form & PTBXL-Sub & PTBXL-Super & CPSC2018 & Average \\
% \midrule
% MERL & 74.4 & 78.5 & 65.9 & 75.7 & 74.2 & 82.8 & 75.3 \\
% D-BETA & 76.3 & 88.6 & 66.1 & 75.9 & 76.2 & 80.1 & 77.2 \\
% MELP & 77.6 & 85.4 & 69.1 & \underline{81.2} & 76.2 & 84.2 & 79.0 \\
% ZETA & 79.2 & 87.5 & \textbf{73.1} & 79.6 & 74.0 & 81.8 & 79.2 \\
% % SCAR & \underline{80.6} & \underline{88.9} & 69.3 & 80.9 & \underline{91.4} & \underline{86.1} & \underline{82.9} \\
% \midrule
% NESA (Ours) & \textbf{82.6} & \textbf{89.2} & \underline{72.1} & \textbf{85.4} & \textbf{96.8} & \textbf{94.4} & \textbf{86.74} \\
% \bottomrule
% \end{tabular}
\begin{tabular}{lccccccc}
\toprule
Methods & CSN & PTBXL-Rhythm & PTBXL-Form & PTBXL-Sub & PTBXL-Super & CPSC2018 & Average \\
\midrule
MERL 
& 74.4 & 78.5 & 65.9 & 75.7 & 74.2 & 82.8 & 75.3 \\

D-BETA 
& 76.3 & \underline{88.6} & 66.1 & 75.9 & \underline{76.2} & 80.1 & 77.2 \\

MELP 
& 77.6 & 85.4 & 69.1 & \underline{81.2} & \underline{76.2} & \underline{84.2} & 79.0 \\

ZETA 
& \underline{79.2} & 87.5 & \textbf{73.1} & 79.6 & 74.0 & 81.8 & \underline{79.2} \\

\midrule

NESA (Ours) 
& \textbf{82.6} 
& \textbf{89.2} 
& \underline{72.1} 
& \textbf{85.4} 
& \textbf{96.8} 
& \textbf{94.4} 
& \textbf{86.7} \\

\bottomrule
\end{tabular}
}
\end{table}

\begin{table}
\centering
\caption{Zero-shot classification performance (AUROC [\%]) using label-only text inputs across multiple datasets. Values in parentheses indicate the performance change compared with prompt-based zero-shot classification.}
\label{tab:main_zeroshot_label}
\resizebox{\columnwidth}{!}{
\begin{tabular}{lccccccc}
\toprule
Methods & CSN & PTBXL-Rhythm & PTBXL-Form & PTBXL-Sub & PTBXL-Super & CPSC2018 & Average \\
\midrule

MERL 
& 62.1 ($\downarrow$12.3) 
& 67.5 ($\downarrow$11.0) 
& 52.8 ($\downarrow$13.1) 
& 72.8 ($\downarrow$2.9) 
& 52.4 ($\downarrow$21.8) 
& 72.0 ($\downarrow$10.8) 
& 63.3 ($\downarrow$12.0) \\

D-BETA
& \underline{68.5} ($\downarrow$7.8)
& \underline{83.6} ($\downarrow$5.0)
& 61.5 ($\downarrow$4.6)
& \underline{73.3} ($\downarrow$2.6)
& \underline{71.6} ($\downarrow$4.6)
& 76.5 ($\downarrow$3.6)
& \underline{72.5} ($\downarrow$4.7) \\

MELP 
& 67.8 ($\downarrow$9.8) 
& 80.7 ($\downarrow$4.7) 
& \underline{62.5} ($\downarrow$6.6) 
& 71.2 ($\downarrow$10.0) 
& 66.1 ($\downarrow$10.1) 
& \textbf{79.8} ($\downarrow$4.4)
& 71.4 ($\downarrow$7.6) \\

ZETA
& \underline{68.5} ($\downarrow$10.7)
& \underline{83.6} ($\downarrow$3.9)
& 61.5 ($\downarrow$11.6)
& \underline{73.3} ($\downarrow$6.3)
& \underline{71.6} ($\downarrow$2.4)
& 76.5 ($\downarrow$5.3)
& \underline{72.5} ($\downarrow$6.7) \\

\midrule

NESA (Ours) 
& \textbf{69.3} ($\downarrow$13.3) 
& \textbf{90.1} ($\uparrow$0.9) 
& \textbf{72.5} ($\uparrow$0.4) 
& \textbf{80.7} ($\downarrow$4.7) 
& \textbf{86.9} ($\downarrow$9.9) 
& \underline{78.6} ($\downarrow$15.8)
& \textbf{79.7} ($\downarrow$7.0) \\

\bottomrule
\end{tabular}
}
\end{table}

\begin{table*}[t]
\centering
\caption{Random-token deletion robustness of NESA and MELP. We randomly remove $k$\% ECGFM backbone tokens at inference time and report the remaining zero-shot AUROC (\%). The AUROC change relative to the full-token setting is shown in parentheses. Changes smaller than 0.01 percentage points are indicated by $\downarrow <0.01$. Average AUROC is computed across the six evaluation tasks. Higher AUROC and smaller performance degradation indicate better robustness to non-targeted token loss.}
\label{tab:random_token_deletion}
\resizebox{\textwidth}{!}{
\begin{tabular}{llccccccc}
\toprule
Deletion Ratio & Methods & CSN & PTBXL-Rhythm & PTBXL-Form & PTBXL-Sub & PTBXL-Super & CPSC2018 & Average \\
\midrule
\multirow{2}{*}{Rand-1\%}
& MELP
& \underline{76.59} ($\downarrow$1.01)
& \underline{84.76} ($\downarrow$0.64)
& \underline{69.05} ($\downarrow$0.05)
& \underline{80.53} ($\downarrow$0.67)
& \underline{76.05} ($\downarrow$0.15)
& \underline{83.89} ($\downarrow$0.31)
& \underline{78.48} ($\downarrow$0.47) \\
& NESA (Ours)
& \textbf{82.61} ($\downarrow <0.01$)
& \textbf{89.15} ($\downarrow <0.01$)
& \textbf{72.13} ($\downarrow <0.01$)
& \textbf{85.36} ($\downarrow <0.01$)
& \textbf{96.78} ($\downarrow <0.01$)
& \textbf{94.43} ($\downarrow <0.01$)
& \textbf{86.74} ($\downarrow <0.01$) \\
\midrule
\multirow{2}{*}{Rand-5\%}
& MELP
& \underline{75.99} ($\downarrow$1.61)
& \underline{84.43} ($\downarrow$0.97)
& \underline{69.01} ($\downarrow$0.09)
& \underline{80.08} ($\downarrow$1.12)
& \underline{75.90} ($\downarrow$0.30)
& \underline{83.80} ($\downarrow$0.40)
& \underline{78.20} ($\downarrow$0.75) \\
& NESA (Ours)
& \textbf{82.61} ($\downarrow <0.01$)
& \textbf{89.15} ($\downarrow <0.01$)
& \textbf{72.13} ($\downarrow <0.01$)
& \textbf{85.36} ($\downarrow <0.01$)
& \textbf{96.78} ($\downarrow <0.01$)
& \textbf{94.43} ($\downarrow <0.01$)
& \textbf{86.74} ($\downarrow <0.01$) \\
\midrule
\multirow{2}{*}{Rand-10\%}
& MELP
& \underline{75.71} ($\downarrow$1.89)
& \underline{84.26} ($\downarrow$1.14)
& \underline{68.96} ($\downarrow$0.14)
& \underline{79.85} ($\downarrow$1.35)
& \underline{75.81} ($\downarrow$0.39)
& \underline{83.80} ($\downarrow$0.40)
& \underline{78.07} ($\downarrow$0.89) \\
& NESA (Ours)
& \textbf{82.61} ($\downarrow <0.01$)
& \textbf{89.15} ($\downarrow <0.01$)
& \textbf{72.13} ($\downarrow <0.01$)
& \textbf{85.36} ($\downarrow <0.01$)
& \textbf{96.78} ($\downarrow <0.01$)
& \textbf{94.43} ($\downarrow <0.01$)
& \textbf{86.74} ($\downarrow <0.01$) \\
\midrule
\multirow{2}{*}{Rand-20\%}
& MELP
& \underline{75.45} ($\downarrow$2.15)
& \underline{84.10} ($\downarrow$1.30)
& \underline{68.95} ($\downarrow$0.15)
& \underline{79.64} ($\downarrow$1.56)
& \underline{75.72} ($\downarrow$0.48)
& \underline{83.81} ($\downarrow$0.39)
& \underline{77.95} ($\downarrow$1.01) \\
& NESA (Ours)
& \textbf{82.60} ($\downarrow <0.01$)
& \textbf{89.15} ($\downarrow <0.01$)
& \textbf{72.13} ($\downarrow <0.01$)
& \textbf{85.36} ($\downarrow <0.01$)
& \textbf{96.78} ($\downarrow <0.01$)
& \textbf{94.43} ($\downarrow <0.01$)
& \textbf{86.74} ($\downarrow <0.01$) \\
\bottomrule
\end{tabular}
}
\end{table*}

\begin{table*}
\centering
\caption{Dominant-token deletion robustness of NESA and MELP. We remove the top-$k$\% most alignment-dominant ECG tokens at inference time and report the remaining zero-shot AUROC (\%). The AUROC drop relative to the full-token setting is shown in parentheses, where $\downarrow <0.01$ indicates a decrease smaller than 0.01 percentage points. Average AUROC is computed across the six evaluation tasks. Higher AUROC and smaller performance degradation indicate better robustness.}
\label{tab:dominant_delete}
\resizebox{\textwidth}{!}{
\begin{tabular}{llccccccc}
\toprule
Deletion Ratio & Methods & CSN & PTBXL-Rhythm & PTBXL-Form & PTBXL-Sub & PTBXL-Super & CPSC2018 & Average \\
\midrule
\multirow{2}{*}{Top-1\%}
& MELP
& \underline{77.21} ($\downarrow$0.39)
& \underline{84.98} ($\downarrow$0.42)
& \underline{68.57} ($\downarrow$0.53)
& \underline{80.46} ($\downarrow$0.74)
& \underline{75.34} ($\downarrow$0.86)
& \underline{83.29} ($\downarrow$0.91)
& \underline{78.31} ($\downarrow$0.64) \\
& NESA (Ours)
& \textbf{82.61} ($\downarrow <0.01$)
& \textbf{89.15} ($\downarrow <0.01$)
& \textbf{72.13} ($\downarrow <0.01$)
& \textbf{85.36} ($\downarrow <0.01$)
& \textbf{96.78} ($\downarrow <0.01$)
& \textbf{94.43} ($\downarrow <0.01$)
& \textbf{86.74} ($\downarrow <0.01$) \\

\midrule
\multirow{2}{*}{Top-5\%}
& MELP
& \underline{75.99} ($\downarrow$1.61)
& \underline{84.65} ($\downarrow$0.75)
& \underline{68.22} ($\downarrow$0.88)
& \underline{79.95} ($\downarrow$1.25)
& \underline{74.32} ($\downarrow$1.88)
& \underline{79.42} ($\downarrow$4.78)
& \underline{77.09} ($\downarrow$1.86) \\
& NESA (Ours)
& \textbf{82.61} ($\downarrow <0.01$)
& \textbf{89.15} ($\downarrow <0.01$)
& \textbf{72.13} ($\downarrow <0.01$)
& \textbf{85.36} ($\downarrow <0.01$)
& \textbf{96.78} ($\downarrow <0.01$)
& \textbf{94.43} ($\downarrow <0.01$)
& \textbf{86.74} ($\downarrow <0.01$) \\

\midrule
\multirow{2}{*}{Top-10\%}
& MELP
& \underline{74.15} ($\downarrow$3.45)
& \underline{81.87} ($\downarrow$3.53)
& \underline{68.08} ($\downarrow$1.02)
& \underline{78.52} ($\downarrow$2.68)
& \underline{69.85} ($\downarrow$6.35)
& \underline{74.66} ($\downarrow$9.54)
& \underline{74.52} ($\downarrow$4.43) \\
& NESA (Ours)
& \textbf{82.61} ($\downarrow <0.01$)
& \textbf{89.15} ($\downarrow <0.01$)
& \textbf{72.13} ($\downarrow <0.01$)
& \textbf{85.36} ($\downarrow <0.01$)
& \textbf{96.78} ($\downarrow <0.01$)
& \textbf{94.43} ($\downarrow <0.01$)
& \textbf{86.74} ($\downarrow <0.01$) \\

\midrule
\multirow{2}{*}{Top-20\%}
& MELP
& \underline{72.94} ($\downarrow$4.66)
& \underline{78.02} ($\downarrow$7.38)
& \underline{63.23} ($\downarrow$5.87)
& \underline{75.58} ($\downarrow$5.62)
& \underline{63.76} ($\downarrow$12.44)
& \underline{73.52} ($\downarrow$10.68)
& \underline{71.18} ($\downarrow$7.78) \\
& NESA (Ours)
& \textbf{82.52} ($\downarrow$0.09)
& \textbf{89.15} ($\downarrow <0.01$)
& \textbf{72.08} ($\downarrow$0.05)
& \textbf{85.34} ($\downarrow$0.02)
& \textbf{96.58} ($\downarrow$0.20)
& \textbf{94.41} ($\downarrow$0.02)
& \textbf{86.68} ($\downarrow$0.06) \\
\bottomrule
\end{tabular}
}
\end{table*}

% \subsection{Token Deletion Robustness Analysis}
% \label{sec:dominant_token}

% We evaluate token-level robustness using two complementary inference-time deletion protocols: random token deletion and dominant-token deletion. Random deletion uniformly removes $k\%$ ECGFM backbone tokens and serves as a control for non-targeted token loss, while dominant-token deletion removes the top-$k\%$ most alignment-dominant ECG tokens to directly test whether a model relies on a small subset of shortcut evidence. Both protocols are conducted at inference time without parameter updates. For dominant-token deletion, token ranking is used only for robustness analysis and does not provide labels or paired reports to NESA inference.

% Table~\ref{tab:random_token_deletion} reports the random deletion results. Since the removed tokens are sampled uniformly, this protocol mainly measures general robustness to token-level perturbations. Across deletion ratios from 1\% to 20\%, NESA maintains nearly unchanged zero-shot AUROC, while MELP and SCAR also show only mild degradation. This indicates that non-targeted token loss is not sufficient to expose dominant-token shortcut reliance, and therefore serves mainly as a robustness control.
\subsection{Token Deletion Robustness Analysis}
\label{sec:dominant_token}

We evaluate token-level robustness using two complementary inference-time deletion protocols. Random deletion uniformly removes $k\%$ ECGFM backbone tokens and serves as a control for non-targeted token loss, whereas dominant-token deletion removes the top-$k\%$ most alignment-dominant tokens to test reliance on a small subset of alignment-critical evidence.

Table~\ref{tab:random_token_deletion} reports the random deletion results. Across deletion ratios from 1\% to 20\%, NESA maintains an average zero-shot AUROC of 86.74\%, whereas MELP decreases from 78.48\% to 77.95\%. The small changes under non-targeted deletion provide a control for interpreting the dominant-token deletion results.

\begin{figure}
\centering
\includegraphics[width=\textwidth]{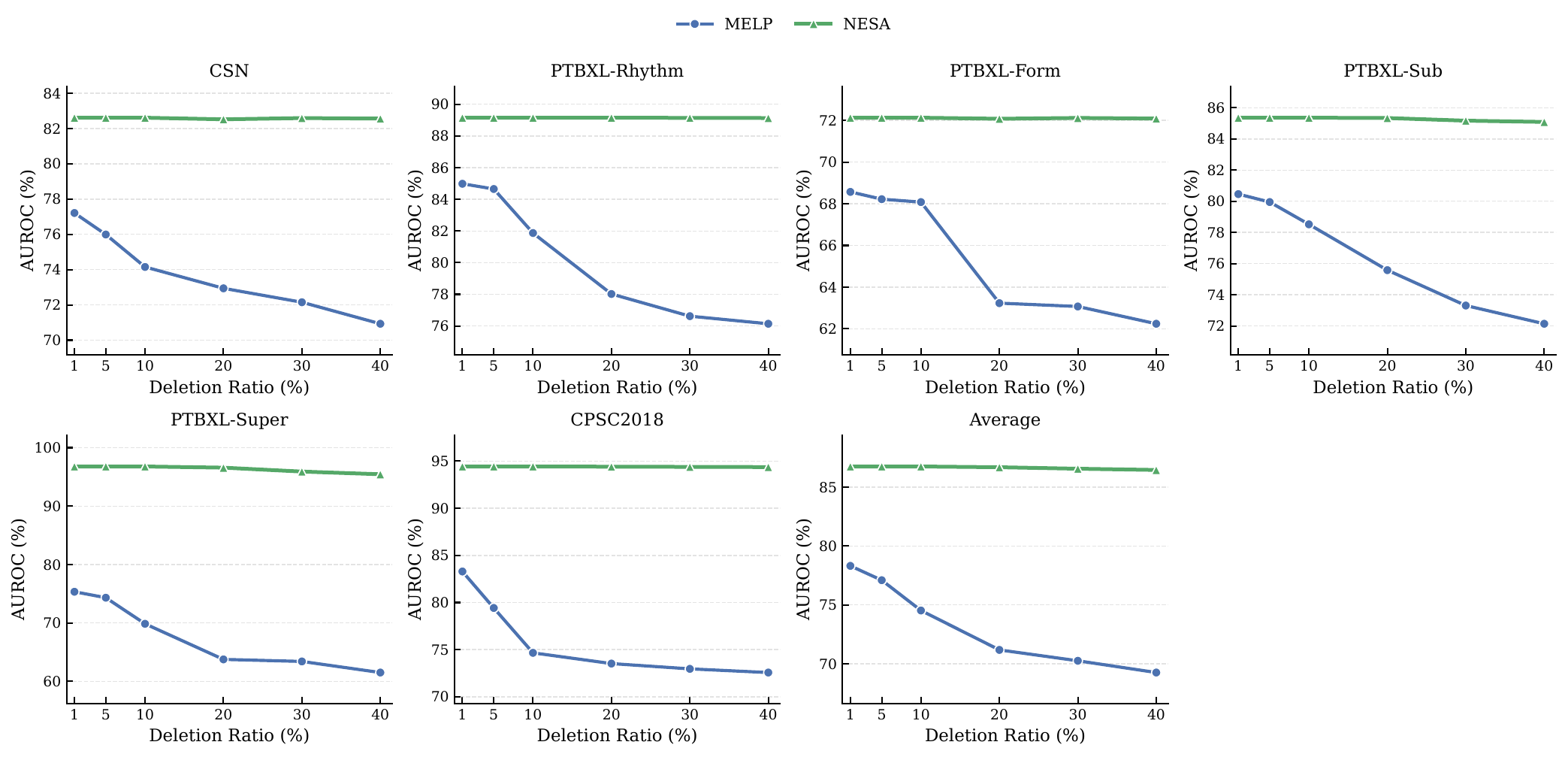}
% \caption{Dominant-token deletion curves across datasets. We progressively remove the top-ranked alignment-dominant ECG tokens and evaluate the remaining zero-shot AUROC on CSN, PTB-XL subsets, CPSC2018, and the averaged result. MELP and SCAR exhibit clear performance degradation as more dominant tokens are removed, indicating their dependence on a small subset of highly aligned evidence. In contrast, NESA maintains stable AUROC across deletion ratios, suggesting that the learned ECG representation relies on more distributed and complementary diagnostic evidence.}
\caption{Dominant-token deletion curves across six evaluation tasks and their average. We progressively remove the top-ranked alignment-dominant ECG tokens at inference time and evaluate the remaining zero-shot AUROC under deletion ratios ranging from 1\% to 40\%. MELP exhibits substantial performance degradation as more dominant tokens are removed, whereas NESA maintains relatively stable AUROC across deletion ratios. These results suggest that NESA is less dependent on a small subset of highly aligned tokens, indicating greater robustness to the removal of dominant ECG evidence.}
\label{fig:dominant_deletion_curves}
\end{figure}

Table~\ref{tab:dominant_delete} and Fig.~\ref{fig:dominant_deletion_curves} report dominant-token deletion results. When the top 20\% dominant tokens are removed, MELP loses 7.78 average AUROC points, whereas NESA loses 0.06 points and shows negligible degradation up to 10\% deletion. Together with the random-deletion control, this result indicates lower dependence on a small subset of highly aligned ECG tokens.

% Table~\ref{tab:dominant_delete} and Fig.~\ref{fig:dominant_deletion_curves} further report dominant-token deletion results, where the most alignment-dominant ECG tokens are removed in the latent token space used by the shared aggregation encoder. This protocol reveals a much clearer difference among methods. When the top 20\% dominant tokens are removed, MELP and SCAR drop by 7.78 and 3.44 average AUROC points, respectively, whereas NESA drops by only 0.06 points and remains almost unchanged even under 10\% deletion. These results suggest that NESA does not solve ECG--text matching through a few highly salient tokens, but instead grounds textual semantics in more distributed and complementary ECG evidence.

\subsection{Evidence Coverage and Qualitative Redistribution}
\label{sec:evidence_coverage}

% \begin{table*}
% \centering
% \caption{Evidence coverage statistics during inference. Top-$5$\% concentration measures the proportion of total token importance mass assigned to the top $5$\% most important ECG tokens. Entropy is computed over the normalized token importance distribution. Lower concentration and higher entropy indicate broader evidence usage.}
% \label{tab:coverage_stats}
% \resizebox{\textwidth}{!}{
% \begin{tabular}{llccccccc}
% \toprule
% Metric & Method
% & CSN & PTBXL-Rhythm & PTBXL-Form & PTBXL-Sub & PTBXL-Super & CPSC2018 & Average \\
% \midrule
% \multirow{3}{*}{Top-5\% Conc. $\downarrow$}
% & MELP & 34.90 & 51.63 & 25.74 & 37.86 & 28.31 & 47.24 & 37.61 \\
% & SCAR & \underline{33.43} & \underline{35.56} & \underline{25.67} & \underline{29.42} & \underline{24.34} & \underline{42.40} & \underline{31.80} \\
% & NESA (Ours) & \textbf{5.41} & \textbf{6.45} & \textbf{5.85} & \textbf{8.07} & \textbf{5.43} & \textbf{8.11} & \textbf{6.55} \\
% \midrule
% \multirow{3}{*}{Entropy $\uparrow$}
% & MELP & 4.39 & 3.74 & \underline{4.77} & 4.06 & 4.47 & 3.83 & 4.21 \\
% & SCAR & \underline{4.43} & \underline{4.43} & 4.62 & \underline{4.65} & \underline{4.83} & \underline{3.91} & \underline{4.48} \\
% & NESA (Ours) & \textbf{5.73} & \textbf{5.68} & \textbf{5.71} & \textbf{5.61} & \textbf{5.73} & \textbf{5.59} & \textbf{5.68} \\
% \bottomrule
% \end{tabular}
% }
% \end{table*}
\begin{table*}
\centering
% \caption{Evidence coverage statistics during inference. Top-5\% concentration measures the percentage of total normalized importance mass assigned to the top 5\% most important ECG tokens. Token importance scores are normalized to sum to one, $p_i = w_i / \sum_{j=1}^{N} w_j$, where $N=1024$. Entropy is computed as $H=-\sum_{i=1}^{N}p_i\log p_i$. Lower concentration and higher entropy indicate a less concentrated importance distribution.}
\caption{Evidence coverage statistics during inference. Top-5\% concentration denotes the percentage of total normalized token-importance mass assigned to the top 5\% most important ECG tokens. Entropy measures the dispersion of the normalized token-importance distribution. Lower concentration and higher entropy indicate broader evidence usage.}
\label{tab:coverage_stats}
\resizebox{\textwidth}{!}{
\begin{tabular}{llccccccc}
\toprule
Metric & Method
& CSN & PTBXL-Rhythm & PTBXL-Form & PTBXL-Sub & PTBXL-Super & CPSC2018 & Average \\
\midrule
\multirow{2}{*}{Top-5\% Conc. $\downarrow$}
& MELP
& \underline{34.90}
& \underline{51.63}
& \underline{25.74}
& \underline{37.86}
& \underline{28.31}
& \underline{47.24}
& \underline{37.61} \\
& NESA (Ours)
& \textbf{5.41}
& \textbf{6.45}
& \textbf{5.85}
& \textbf{8.07}
& \textbf{5.43}
& \textbf{8.11}
& \textbf{6.55} \\
\midrule
\multirow{2}{*}{Entropy $\uparrow$}
& MELP
& \underline{4.39}
& \underline{3.74}
& \underline{4.77}
& \underline{4.06}
& \underline{4.47}
& \underline{3.83}
& \underline{4.21} \\
& NESA (Ours)
& \textbf{5.73}
& \textbf{5.68}
& \textbf{5.71}
& \textbf{5.61}
& \textbf{5.73}
& \textbf{5.59}
& \textbf{5.68} \\
\bottomrule
\end{tabular}
}
\end{table*}

\begin{figure}
\centering
\includegraphics[width=\textwidth]{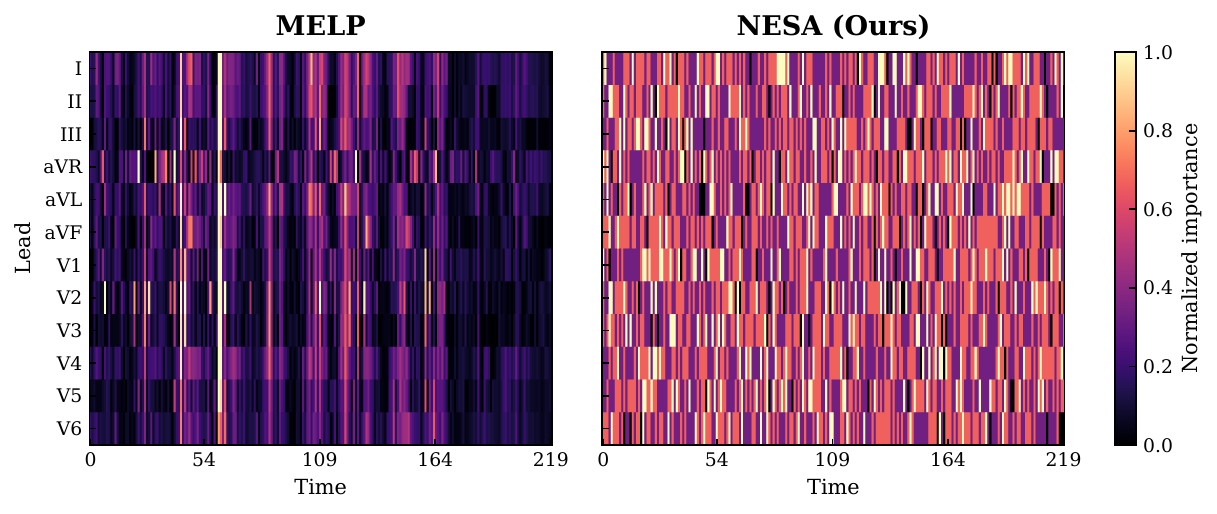}
% \caption{Spatial-temporal token importance maps of different ECG--text pretraining methods. Each heatmap visualizes the normalized text-aligned importance over ECG leads and temporal tokens. MELP concentrates high importance on a limited set of leads and time segments, while SCAR produces sparse and irregular evidence patterns after adversarial masking. NESA yields a more globally distributed importance map across both limb and precordial leads, indicating that progressive counterfactual erasing encourages the model to exploit broader ECG evidence.}
\caption{Spatial-temporal token importance maps of MELP and NESA. Each heatmap visualizes normalized text-aligned importance over ECG leads and temporal tokens. MELP concentrates high importance on a limited set of leads and time segments, whereas NESA exhibits a broader pattern across both limb and precordial leads.}
\label{fig:token_importance_map}
\end{figure}

\begin{figure}
\centering
\includegraphics[width=\textwidth]{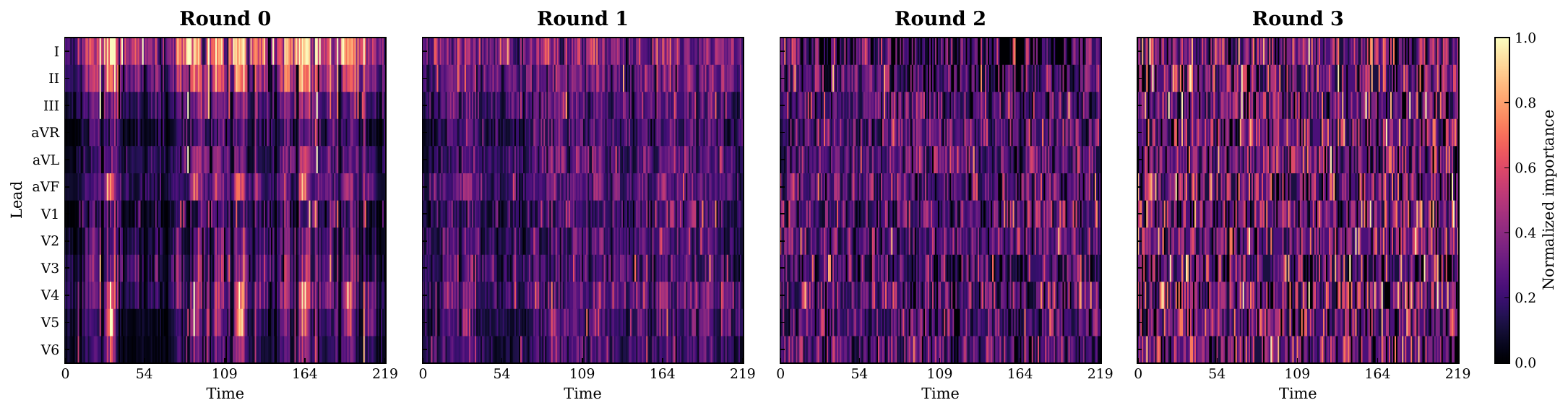}
\caption{Evolution of token importance during progressive erasing. The heatmaps show the normalized ECG token importance from the original view to successive erasing rounds. In early rounds, NESA identifies highly text-aligned evidence and suppresses these dominant regions. As the erasing process proceeds, the model is forced to activate residual diagnostic cues from previously underused leads and temporal segments. This progressive redistribution demonstrates how NESA expands evidence coverage and mitigates shortcut alignment to a fixed set of salient ECG tokens.}
\label{fig:progressive_round_importance}
\end{figure}
\begin{figure}
\centering
\includegraphics[width=\columnwidth]{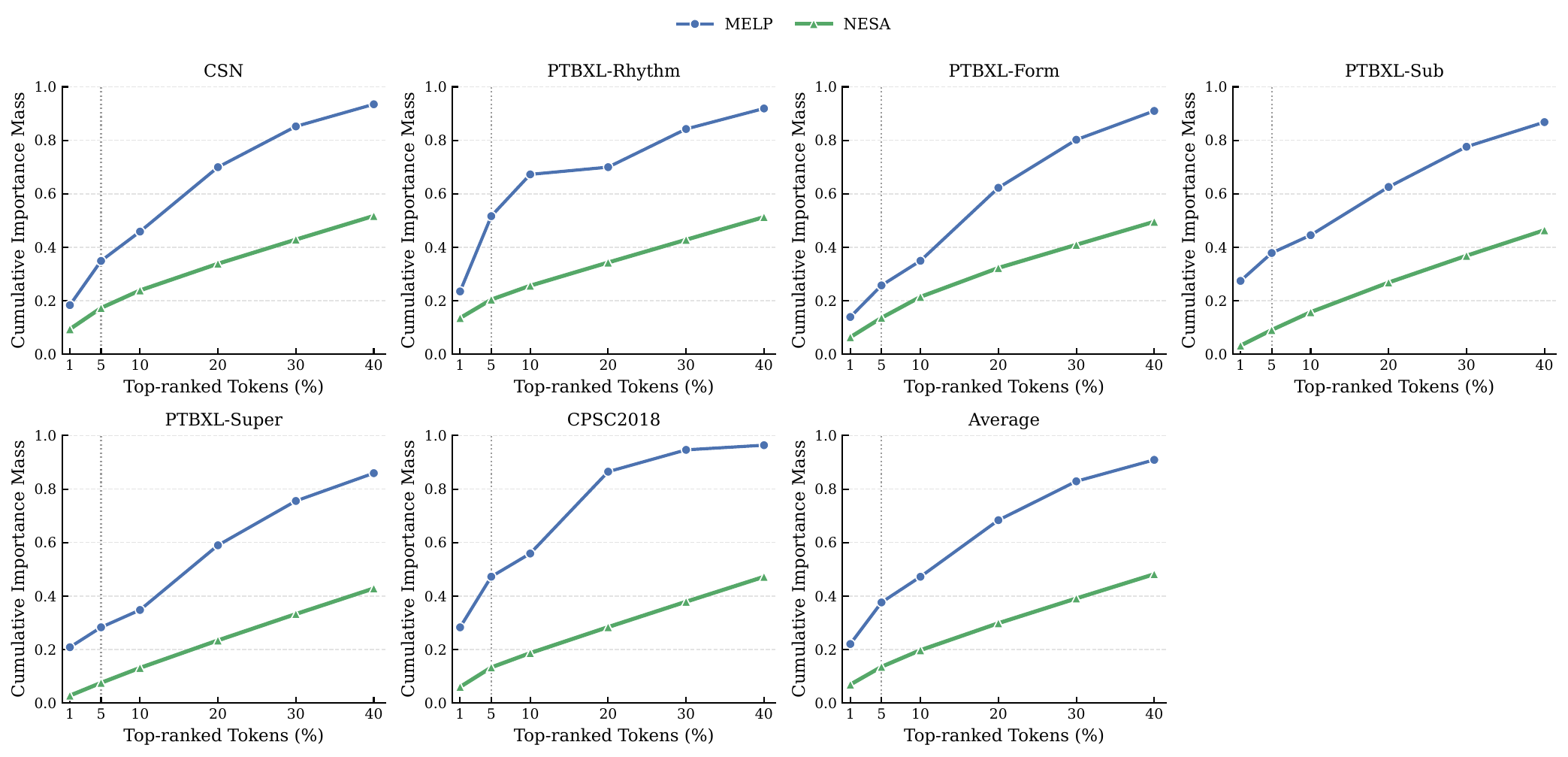}
% \caption{Cumulative importance mass of top-ranked ECG tokens across datasets. 
% The x-axis denotes the proportion of top-ranked alignment-dominant ECG tokens, and the y-axis denotes the cumulative importance mass assigned to these tokens. A steeper curve indicates that the model concentrates more importance on a small subset of dominant tokens, whereas a flatter curve indicates broader evidence usage. Across all datasets, MELP and SCAR accumulate importance mass more rapidly among top-ranked tokens, while NESA shows a consistently flatter profile, indicating less dominant-token reliance and more distributed ECG--text grounding.}
\caption{Cumulative importance mass of top-ranked ECG tokens across datasets. The x-axis denotes the proportion of top-ranked alignment-dominant ECG tokens, and the y-axis denotes the cumulative importance mass assigned to these tokens. A steeper curve indicates that the model concentrates more importance on a small subset of dominant tokens, whereas a flatter curve indicates broader evidence coverage. Across all datasets, MELP accumulates importance mass more rapidly among top-ranked tokens, while NESA shows a consistently flatter profile, indicating lower importance concentration and reduced reliance on dominant ECG tokens.}
\label{fig:cumulative_importance_mass}
\end{figure}

% We next examine whether the gain of NESA indeed comes from broader evidence usage. Table~\ref{tab:coverage_stats} reports quantitative coverage statistics, while Figs.~\ref{fig:token_importance_map}, \ref{fig:progressive_round_importance}, and \ref{fig:cumulative_importance_mass} provide qualitative and distributional evidence.

% Quantitatively, NESA substantially reduces evidence concentration. Its average Top-5\% concentration is 6.55, compared with 31.80 for SCAR and 37.61 for MELP, indicating that much less of the total importance mass is monopolized by a small set of highly ranked tokens. At the same time, NESA achieves the highest average token-importance entropy of 5.68, compared with 4.48 for SCAR and 4.21 for MELP, showing that the learned importance distribution is substantially flatter and broader.
We next examine whether NESA's robustness is accompanied by broader evidence usage. Table~\ref{tab:coverage_stats} reports quantitative coverage statistics, while Figs.~\ref{fig:token_importance_map}, \ref{fig:progressive_round_importance}, and \ref{fig:cumulative_importance_mass} provide qualitative and distributional evidence.

NESA lowers the average Top-5\% concentration from 37.61\% for MELP to 6.55\% and raises the average token-importance entropy from 4.21 to 5.68. Together with the token-deletion results, these statistics indicate a broader importance distribution and reduced reliance on dominant ECG tokens.

% These quantitative findings are consistent with the visualizations. In Fig.~\ref{fig:token_importance_map}, MELP and SCAR concentrate importance on limited lead--time regions, whereas NESA distributes importance over a wider range of limb and precordial leads. Fig.~\ref{fig:progressive_round_importance} further shows that, as dominant tokens are progressively erased, the model shifts toward previously underused diagnostic regions instead of repeatedly collapsing onto a fixed subset. Finally, Fig.~\ref{fig:cumulative_importance_mass} shows that NESA accumulates importance more slowly among top-ranked tokens, providing additional evidence that it alleviates dominant-token reliance and encourages distributed ECG--text grounding.
These quantitative findings are consistent with the visualizations. In Fig.~\ref{fig:token_importance_map}, MELP concentrates importance on limited lead--time regions, whereas NESA distributes importance over a wider range of limb and precordial leads. Fig.~\ref{fig:progressive_round_importance} further illustrates how the importance distribution evolves across successive evidence erasing rounds, with alignment-relevant evidence shifting toward previously less emphasized ECG regions. Finally, Fig.~\ref{fig:cumulative_importance_mass} shows that NESA accumulates importance more slowly among top-ranked tokens than MELP, providing additional evidence that it alleviates dominant-token reliance and promotes broader ECG evidence coverage.

\begin{table*}
\centering
\caption{Ablation study of the proposed framework. We report zero-shot AUROC (\%) and Top-5\% concentration. Lower concentration indicates broader evidence usage.}
\label{tab:ablation}
\resizebox{\textwidth}{!}{
\begin{tabular}{llccccccc}
\toprule
Metric & Variant
& CSN & PTBXL-Rhythm & PTBXL-Form & PTBXL-Sub & PTBXL-Super & CPSC2018 & Average \\
\midrule
\multirow{6}{*}{AUROC $\uparrow$}
& Full model
& \textbf{82.61} & 89.15 & \textbf{72.13} & \textbf{85.36} & \textbf{96.78} & \textbf{94.43} & \textbf{86.74} \\
& w/o round-wise alignment correction
& \underline{76.29} & \textbf{90.92} & \underline{70.51} & \underline{82.32} & \underline{88.83} & 87.89 & \underline{82.79} \\
& w/o progressive erasing
& 71.15 & 87.85 & 64.83 & 76.08 & 83.39 & 83.42 & 77.79 \\
& w/o alignment-aware scoring
& 75.10 & 89.67 & 69.59 & 77.23 & 87.97 & \underline{88.74} & 81.38 \\
& w/o semantic consistency
& 76.03 & 89.41 & 71.48 & 81.38 & 87.54 & 85.56 & 81.90 \\
& w/o coverage regularization
& 74.37 & \underline{90.79} & 68.71 & 81.96 & 88.61 & 88.14 & 82.10 \\
\midrule
\multirow{6}{*}{Top-5\% Conc. $\downarrow$}
& Full model
& \textbf{5.41} & \textbf{6.45} & \textbf{5.85} & \underline{8.07} & \textbf{5.43} & \textbf{8.11} & \textbf{6.55} \\
& w/o round-wise alignment correction
& \underline{7.96} & 9.01 & \underline{7.64} & \textbf{7.81} & \underline{7.88} & \underline{8.52} & \underline{8.14} \\
& w/o progressive erasing
& 9.47 & 9.69 & 9.04 & 10.04 & 9.93 & 15.47 & 10.61 \\
& w/o alignment-aware scoring
& 8.44 & 9.67 & 8.44 & 8.30 & 8.28 & 16.21 & 9.89 \\
& w/o semantic consistency
& 10.01 & 10.46 & 9.38 & 8.11 & 8.26 & 17.14 & 10.56 \\
& w/o coverage regularization
& 8.50 & \underline{8.89} & 8.70 & 8.44 & 8.65 & 11.20 & 9.06 \\
\bottomrule
\end{tabular}
}
\end{table*}

\subsection{Ablation Study}
\label{sec:ablation}

Table~\ref{tab:ablation} reports ablation results for the major components of NESA. Removing progressive erasing causes the largest performance drop: the average zero-shot AUROC decreases from 86.74\% to 77.79\%, while the average Top-5\% concentration increases from 6.55 to 10.61. This confirms that repeatedly suppressing dominant ECG evidence is the central driver for breaking shortcut reliance. Furthermore, removing round-wise alignment correction decreases the average AUROC to 82.79\%; this substantial drop highlights the necessity of dynamically updating the cross-modal anchor to prevent semantic collapse when the evidence distribution shifts.

Removing alignment-aware scoring also substantially weakens the model, reducing the average AUROC to 81.38\% and increasing the average concentration to 9.89, which shows that effective erasing depends on dynamically identifying alignment-critical ECG tokens based on the dynamic alignment anchor. Removing semantic consistency lowers the average AUROC to 81.90\% and increases the average Top-5\% concentration from 6.55 to 10.56, indicating that broader evidence usage alone is insufficient unless the erased views remain diagnostically consistent with the original ECG and paired text. Finally, removing coverage regularization leads to a clear drop in average AUROC, from 86.74\% to 82.10\%, together with a concentration increase from 6.55 to 9.06. This shows that progressive erasing benefits from explicitly discouraging repeated reuse of the same evidence pattern across rounds.

\subsection{Hyperparameter Sensitivity Analysis}
\label{sec:hyperparameter}

We further analyze the sensitivity of NESA to the grouped auxiliary loss coefficients, the number of erasing rounds, and the erasing ratio per round. Unless otherwise specified, we vary one hyperparameter group at a time while keeping the others fixed to the default setting. All results are averaged over the zero-shot evaluation subsets.

\paragraph{Effect of grouped loss coefficients.}
Table~\ref{tab:hyper_loss} analyzes the three grouped auxiliary coefficients: the robust alignment coefficient $\alpha_{\text{align}}$, the semantic-preserving consistency coefficient $\beta_{\text{sem/cons}}$, and the restoration-diversity coefficient $\gamma_{\text{rest/div}}$. These coefficients act as group-level multipliers, while the individual loss weights within each group are fixed to their default values.
Setting all three coefficients to 1.0 yields the highest average AUROC (86.74\%), the lowest Top-5\% concentration (6.55), and the highest entropy (5.68). Reducing $\alpha_{\text{align}}$ to 0.5 lowers the average AUROC to 82.46\%, while increasing it to 2.0 further decreases AUROC to 81.78\%, indicating that overly weak or overly strong robust alignment supervision can both hurt ECG--text grounding. Reducing $\beta_{\text{sem/cons}}$ to 0.5 decreases AUROC to 83.78\%, whereas increasing it to 2.0 yields a competitive AUROC of 85.81\% with similar evidence coverage. Finally, changing $\gamma_{\text{rest/div}}$ to 0.5 or 2.0 results in AUROC scores of 86.51\% and 85.40\%, respectively, showing that the restoration-diversity group is relatively stable around the default setting. Overall, the balanced configuration $(1.0,1.0,1.0)$ provides the best trade-off between zero-shot performance and distributed evidence usage.

% \paragraph{Effect of erasing rounds.}
% Table~\ref{tab:hyper_rounds} studies the number of progressive erasing rounds. Increasing the number of rounds from 1 to 3 steadily improves the average AUROC from 81.7\% to 86.7\%, while reducing Top-5\% concentration from 8.9 to 6.6 and increasing entropy from 5.56 to 5.68. This confirms that multi-round erasing is beneficial for suppressing dominant-token reliance and exposing complementary evidence. When the number of rounds is further increased to 4, the average AUROC drops to 83.9\%, although the evidence distribution remains relatively broad with a Top-5\% concentration of 7.6. Increasing the rounds to 5 further decreases AUROC to 79.5\%, suggesting that overly deep erasing can remove too much useful diagnostic evidence and destabilize alignment. We therefore use three rounds as the default setting.
\paragraph{Effect of erasing rounds.}
Table~\ref{tab:hyper_rounds} studies the number of progressive erasing rounds. Starting from a single erasing round ($K=1$), increasing the number of rounds to three steadily improves the average AUROC from 81.7\% to 86.7\%, while reducing Top-5\% concentration from 8.9 to 6.6 and increasing entropy from 5.56 to 5.68. These results demonstrate the benefits of multi-round erasing in reducing dominant-token reliance and encouraging complementary evidence utilization. When the number of rounds is further increased to 4, the average AUROC drops to 83.9\%, although the evidence distribution remains relatively broad with a Top-5\% concentration of 7.6. Increasing the rounds to 5 further decreases AUROC to 79.5\%, suggesting that overly deep erasing can remove too much useful diagnostic evidence and destabilize alignment. We therefore use three rounds as the default setting.

\paragraph{Effect of erasing ratio.}
Table~\ref{tab:hyper_ratio} analyzes the erasing ratio per round. A very small ratio of 1\% provides only weak pressure against dominant-token reliance and yields an average AUROC of 83.4\%, with a Top-5\% concentration of 11.1 and entropy of 5.08. Increasing the ratio to 5\% gives the best average AUROC of 86.7\%, while achieving the lowest Top-5\% concentration of 6.6 and the highest entropy of 5.68. A larger ratio of 10\% remains competitive, with an AUROC of 85.6\%, Top-5\% concentration of 7.2, and entropy of 5.66. However, when the ratio is increased to 20\%, the average AUROC drops substantially to 79.5\%, indicating that overly aggressive erasing removes too much diagnostic information. Therefore, 5\% provides the best trade-off between suppressing shortcut evidence and preserving sufficient residual ECG semantics.

\begin{table}
\centering
\caption{
Sensitivity to grouped auxiliary loss coefficients.
$\alpha_{\text{align}}$ controls the robust ECG-text alignment group,
$\beta_{\text{sem/cons}}$ controls the semantic-preserving consistency group,
and $\gamma_{\text{rest/div}}$ controls the compensatory restoration and evidence-diversity group.
Results are averaged over all zero-shot evaluation subsets.
}
\label{tab:hyper_loss}
% \resizebox{\columnwidth}{!}{
\begin{tabular}{ccc|ccc}
\toprule
$\alpha_{\text{align}}$ & $\beta_{\text{sem/cons}}$ & $\gamma_{\text{rest/div}}$
& AUROC $\uparrow$ & Top-5\% Conc. $\downarrow$ & Entropy $\uparrow$ \\
\midrule
1.0 & 1.0 & 1.0 & \textbf{86.74} & \textbf{6.55} & \textbf{5.68} \\
0.5 & 1.0 & 1.0 & 82.46 & 8.35 & 5.55 \\
2.0 & 1.0 & 1.0 & 81.78 & 8.29 & 5.57 \\
1.0 & 0.5 & 1.0 & 83.78 & 7.92 & 5.63 \\
1.0 & 2.0 & 1.0 & 85.81 & \underline{6.66} & \underline{5.67} \\
1.0 & 1.0 & 0.5 & \underline{86.51} & 7.62 & 5.64 \\
1.0 & 1.0 & 2.0 & 85.40 & 7.69 & 5.64 \\
\bottomrule
\end{tabular}
% }
\end{table}

\begin{table}
\centering
\small
\caption{Sensitivity to the number of progressive erasing rounds. Training cost is measured by training time per epoch.}
\label{tab:hyper_rounds}
% \resizebox{\columnwidth}{!}{
\begin{tabular}{ccccc}
\toprule
Rounds $K$ & AUROC $\uparrow$ & Top-5\% Conc. $\downarrow$ & Entropy $\uparrow$ & Train / Epoch (min) $\downarrow$ \\
\midrule
1 & 81.7 & 8.9 & 5.56 & \textbf{247.2} \\
2 & \underline{84.5} & 8.1 & \underline{5.59} & \underline{249.8} \\
3 & \textbf{86.7} & \textbf{6.6} & \textbf{5.68} & 252.4 \\
4 & 83.9 & \underline{7.6} & 5.55 & 255.1 \\
5 & 79.5 & 8.8 & 5.50 & 257.2 \\
\bottomrule
\end{tabular}
% }
\end{table}

\begin{table}
\centering
\caption{Sensitivity to the erasing ratio per round. The erasing ratio controls the proportion of currently available tokens removed at each progressive erasing step.}
\label{tab:hyper_ratio}
% \resizebox{\columnwidth}{!}{
\begin{tabular}{cccc}
\toprule
Erasing Ratio $r$ & AUROC $\uparrow$ & Top-5\% Conc. $\downarrow$ & Entropy $\uparrow$ \\
\midrule
1\%  & 83.4 & 11.1 & 5.08 \\
5\%  & \textbf{86.7} & \textbf{6.6} & \textbf{5.68} \\
10\% & \underline{85.6} & \underline{7.2} & \underline{5.66} \\
20\% & 79.5 & 9.1 & 5.36 \\
\bottomrule
\end{tabular}
% }
\end{table}

\begin{table}
\centering
\small
\caption{Efficiency comparison of NESA and ECG--text pretraining baselines in terms of model size, computational cost, training time, and inference latency.}
\label{tab:efficiency}
\begin{tabular}{lcccc}
\toprule
Method & Params (M) & GFLOPs & Train / Epoch (min) & Infer / Sample (ms) \\
\midrule
MERL & 71.74 & 0.08 & 217.31 & 0.12 \\
MELP & 244.17 & 35.71 & 235.33 & 1.31 \\
\midrule
NESA (Ours) & 249.46 & 37.39 & 252.37 & 1.41 \\
\bottomrule
\end{tabular}
\end{table}

% \subsection{Efficiency Analysis}
% \label{sec:efficiency}

% Table~\ref{tab:efficiency} compares NESA with representative ECG--text baselines in terms of model size, training cost, and inference latency. Relative to SCAR, NESA has nearly identical model size (249.46M vs.\ 248.64M parameters) and the same FLOPs (37.39), while increasing training time per epoch only modestly from 246.75 to 252.37 minutes. For zero-shot evaluation, NESA uses the robust fused-view inference path, which introduces multi-round token scoring and aggregation. Nevertheless, its inference latency remains comparable to SCAR and MELP, achieving 1.31 ms per sample. These results suggest that NESA improves ECG--text grounding without introducing substantial computational overhead.
\subsection{Efficiency Analysis}
\label{sec:efficiency}

Table~\ref{tab:efficiency} compares NESA with ECG--text baselines in terms of model size, computational cost, training time, and inference latency. NESA contains 249.46M parameters and requires 37.39 GFLOPs, with a training time of 252.37 minutes per epoch. Its robust fused-view inference path performs multi-round token scoring, aggregation, and alignment correction, resulting in a latency of 1.41 ms per sample versus 1.31 ms for MELP. NESA therefore adds modest computational cost relative to MELP.

% \section{Conclusion}

% This paper presented NESA, a progressive alignment-aware evidence refinement framework for ECG--text pretraining. NESA addresses the supervision imbalance between sparse clinical text and rich ECG evidence by progressively suppressing alignment-dominant tokens and refining residual diagnostic cues across iterative rounds. Experiments on PTB-XL, CPSC2018, and CSN demonstrate that NESA improves both zero-shot classification and label-efficient transfer over existing ECG--text pretraining baselines. Further dominant-token deletion analysis, evidence coverage statistics, and qualitative visualizations show that NESA reduces shortcut reliance on a few salient ECG tokens and promotes more distributed ECG--text grounding. These results suggest that robust ECG--text pretraining requires not only stronger cross-modal alignment, but also explicit control over how diagnostic evidence is distributed and preserved under sparse textual supervision. Future work will further incorporate expert-annotated ECG evidence regions to assess whether the learned distributed grounding aligns with clinically interpretable diagnostic patterns.
\section{Conclusion}

This paper presented NESA, a progressive alignment-aware evidence refinement framework for ECG--text pretraining. NESA addresses the supervision imbalance between sparse clinical text and rich ECG evidence by progressively suppressing alignment-dominant tokens and refining residual diagnostic cues across iterative rounds. Experiments on PTB-XL, CPSC2018, and CSN show that NESA improves zero-shot classification and label-efficient transfer relative to the evaluated baselines. Dominant-token deletion analysis, evidence coverage statistics, and qualitative visualizations further show reduced reliance on salient ECG tokens and more distributed ECG--text grounding. These results suggest that robust ECG--text pretraining requires both stronger cross-modal alignment and explicit control over how diagnostic evidence is distributed and preserved under sparse textual supervision. Future work will incorporate expert-annotated ECG evidence regions to assess whether the learned distributed grounding aligns with clinically interpretable diagnostic patterns.

\bibliographystyle{IEEEtran}

\bibliography{cas-refs}

\end{document}